\documentclass[preprint,12pt]{elsarticle}

\usepackage{amssymb}
\usepackage{amsthm}
\usepackage{color}
\usepackage{url,hyperref}
\usepackage{lineno}
\usepackage[utf8]{inputenc}
\usepackage[T1]{fontenc}

\journal{Nuclear Inst. and Methods in Physics Research, A}

\begin{document}

\begin{frontmatter}



\title{SIDDHARTA-2 apparatus for kaonic atoms research on the DA$\Phi$NE collider}

\author[inst1,inst2]{F. Sirghi\corref{cor1}} 
\ead{florin.sirghi@lnf.infn.it}
\author[inst1]{F. Sgaramella\corref{cor1}}
\ead{francesco.sgaramella@lnf.infn.it}

\author[inst12]{L. Abbene}
\author[inst5]{C. Amsler}
\author[inst1]{M. Bazzi} 
\author[inst4]{G Borghi}
\author[inst3]{D. Bosnar} 
\author[inst2]{M. Bragadireanu} 
\author[inst12] {A. Buttacavoli} 
\author[inst4]{M. Carminati} 
\author[inst5]{M. Cargnelli} 
\author[inst1]{A. Clozza} 
\author[inst4]{G. Deda} 
\author[inst1]{L. De Paolis} 
\author[inst1,inst6]{R. Del Grande} 
\author[inst1]{K. Dulski} 
\author[inst6]{L. Fabbietti} 
\author[inst4]{C. Fiorini} 
\author[inst3]{I. Fri\v{s}\v{c}i\'c}
\author[inst1]{C. Guaraldo} 
\author[inst1]{M. Iliescu} 
\author[inst9]{M. Iwasaki} 	
\author[inst1,inst7]{A. Khreptak} 
\author[inst1]{S. Manti}
\author[inst5]{J. Marton} 
\author[inst1,inst13]{M. Miliucci} 
\author[inst7,inst8]{P. Moskal} 
\author[inst1]{F. Napolitano} 
\author[inst7,inst8]{S. Nied\'{z}wiecki} 
\author[inst11]{H. Ohnishi} 
\author[inst10,inst1]{K. Piscicchia} 
\author[inst12]{F. Principato}
\author[inst1]{A. Scordo} 
\author[inst10,inst1,inst2]{D. Sirghi} 
\author[inst7,inst8]{M. Skurzok} 
\author[inst7,inst8]{M. Silarski} 
\author[inst1]{A. Spallone} 
\author[inst11]{K. Toho}
\author[inst4]{L. Toscano}
\author[inst5]{M. T\"uchler} 
\author[inst1]{O. Vazquez Doce}
\author[inst11]{C. Yoshida}
\author[inst5]{J. Zmeskal} 
\author[inst1]{C. Curceanu}

\cortext[cor1]{Corresponding author}
 
\affiliation[inst1]{organization={Laboratori Nazionali di Frascati INFN},
            addressline={Via E. Fermi 54}, 
            city={Frascati},
            postcode={00044}, 
            state={Italy}}
\affiliation[inst2]{organization={ Horia Hulubei National Institute of Physics and Nuclear Engineering (IFIN-HH)},
            addressline={Reactorului 30}, 
            city={Magurele},
            postcode={077125}, 
            state={Romania}}
\affiliation[inst3]{organization={Department of Physics, Faculty of Science, University of Zagreb},
            addressline={Bijenicka cesta 32}, 
            city={Zagreb},
            postcode={10000}, 
            state={Croatia}}
\affiliation[inst4]{organization={Politecnico di Milano, Dipartimento di Elettronica, Informazione e Bioingegneria and INFN Sezione di Milano},
            addressline={Via Giuseppe Ponzio 34}, 
            city={Milano},
            postcode={20133}, 
            state={Italy}}
\affiliation[inst5]{organization={Stefan-Meyer-Institut f\"ur Subatomare Physik},
            addressline={Dominikanerbastei 16}, 
            city={Vienna},
            postcode={1010}, 
            state={Austria}}
\affiliation[inst6]{organization={Excellence Cluster Universe, Technische Universiat Munchen Garching},
            addressline={Boltzmann str. 2}, 
            city={Garching},
            postcode={85748}, 
            state={Germany}}

\affiliation[inst7]{organization={Faculty of Physics, Astronomy, and Applied Computer Science, Jagiellonian University},
            addressline={{\L}ojasiewicza 11}, 
            city={Krakow},
            postcode={30-348}, 
            state={Poland}}
            
\affiliation[inst8]{organization={Centre for Theranostics, Jagiellonian University},
addressline={Kopernika 40},
city={Krakow},
postcode={31-501},
state={Poland}}  

\affiliation[inst9]{organization={RIKEN},
            addressline={2-1 Hirosawa, Wako, Saitama}, 
            city={Tokyo},
            postcode={351-0198}, 
            state={Japan}}
            
\affiliation[inst10]{organization={Centro Ricerche Enrico Fermi – Museo Storico della Fisica e Centro Studi e Ricerche “Enrico Fermi”},
            addressline={Via Panisperna 89A}, 
            city={Roma},
            postcode={00184}, 
            state={Italy}}

\affiliation[inst11]{organization={Research Center for Electron Photon Science (ELPH), Tohoku University},
            addressline={1-2-2 Mikamine}, 
            city={Sendai},
            postcode={982-0826}, 
            state={Japan}}

\affiliation[inst12]{organization={Dipartimento di Fisica e Chimica - Emilio Segrè, Università di Palermo},
            addressline={Viale Delle Scienze, Edificio 18}, 
            city={Palermo},
            postcode={90128}, 
            state={ Italy}}

\affiliation[inst13]{organization={ASI, Agenzia Spaziale Italiana},
            addressline={Via del Politecnico snc}, 
            city={Roma},
            postcode={00133}, 
            state={Italy}}

\begin{abstract}

\noindent 

SIDDHARTA-2 represents a state-of-the-art experiment designed to perform dedicated measurements of kaonic atoms, which are particular exotic atom configurations composed of a negatively charged kaon and a nucleus. Investigating these atoms provides an exceptional tool to comprehend the strong interactions in the non-perturbative regime involving strangeness.
The experiment is installed at the DA$\Phi$NE electron-positron collider, of the INFN National Laboratory of Frascati (INFN-LNF) in Italy, aiming to perform the first-ever measurement of the 2p → 1s X-ray transitions in kaonic deuterium, a crucial step towards determining the isospin-dependent antikaon-nucleon scattering lengths. Based on the experience gained with the previous SIDDHARTA experiment, which performed the most precise measurement of the kaonic hydrogen 2p → 1s X-ray transitions, the present apparatus has been upgraded with innovative Silicon Drift Detectors (SDDs), distributed around a cryogenic gaseous target placed in a vacuum chamber at a short distance above the interaction region of the collider.
We present a comprehensive description of the SIDDHARTA-2 setup including the optimization of its various components during the commissioning phase of the collider.

\end{abstract}
\end{frontmatter}

\section{Introduction}\label{introduction}

Light kaonic atoms X-ray spectroscopy is a unique tool for investigating Quantum Chromodynamics (QCD) in the low energy strangeness sector. Precise measurements of the radiative X-ray transitions towards low-n levels in these exotic systems provide information on the kaon-nucleus interaction at threshold which, in typical scattering experiments, would require an extrapolation towards zero energy, making them method-dependent.\\
The electromagnetic interaction of kaons with the nucleus is very well known and the kaonic atoms energy levels can be calculated at a precision of eV by solving the Klein-Gordon equation. Even a small deviation from the electromagnetic value gives information on the strong interaction between the kaon and the nucleus. In the kaonic atoms experiments, the direct observable of interest are the shift ($\epsilon$) and the width ($\Gamma$) of the atomic levels caused by the strong interaction of the kaon with the nucleus \cite{RevModPhys.91.025006}.\\
In this field of research a special role is played by the lightest kaonic atoms, namely kaonic hydrogen, deuterium, and helium. From the first two kaonic atoms, the isospin-dependent antikaon-nucleon scattering lengths can be obtained by measuring the strong interaction-induced shifts and widths of the 1s levels. Additional information on many-body strong interaction can be retrieved from transitions to the 2p level of kaonic helium. The kaonic hydrogen strong interaction induced 1s level shift and width have been measured by the KpX experiment (E228) at KEK \cite{KpX:1997} and by the SIDDHARTA experiment at DA$\Phi$NE in 2009 \cite{Bazzi:2011}, while the first kaonic deuterium measurement is in progress at the DA$\Phi$NE collider by the SIDDHARTA-2 collaboration.\\
To perform the kaonic deuterium measurement, the SIDDHARTA-2 apparatus has been developed to face the challenging very small kaonic deuterium expected X-ray yield \cite{BAZZI201369}. 
We aim to perform the kaonic deuterium measurement with a similar precision as the one of kaonic hydrogen, i.e. precision of approximately 30 eV for the shift and 70 eV for the width. This measurement combined with the existing data from kaonic hydrogen will allow to obtain the antikaon nucleon isospin 0 and 1 scattering lengths. These scattering lengths are fundamental quantities that play a crucial role in enhancing our comprehension of the strong interaction theory within the strangeness sector. Such accuracy is fundamental to efficiently disentangle between different theoretical approaches \cite{Gal:2007,Meisner:2011,Shevchenko:2012,Mizutani:2013,Liu:2020}.\\
Section 2 describes the experimental requirements for kaonic deuterium measurements. In section 3 a detailed description of the experimental apparatus, including the optimization of various components, is introduced. Section 4 presents the results obtained after optimization of the setup for a kaonic helium test run, while conclusions are presented in section 5.

\section{Experimental requirements for the kaonic deuterium measurement}\label{setup_section}

The DA$\Phi$NE accelerator complex is a world-class double ring electron-positron collider operating at the center of mass energy of the $\phi$(1020). The back-to-back $K^+K^-$ pairs, resulting from its almost at-rest decay with a $\simeq 50\%$ branching ratio \cite{Zyla:2020zbs}, have low momentum ($\simeq$ 127~MeV/c) and low energy spread ($\delta p/p\simeq 0.1\%$), making this machine an ideal environment to perform precision spectroscopy of kaonic atoms. DA$\Phi$NE attained its maximum luminosity during the test of the new Crab-Waist collision scheme with the SIDDHARTA apparatus in 2009, achieving a peak luminosity of 4.5$\times$10$^{32}$ cm$^{-2}$s$^{-1}$ \cite{Zobov:2010zza}. 
As already mentioned the experimental challenge for the kaonic deuterium measurement is the very small kaonic deuterium X-ray yield expected to be about one order of magnitude less than for hydrogen, the larger width, and the difficulty to perform precision X-ray spectroscopy in the high radiation environments of the DA$\Phi$NE collider. The collaboration prepared a series of modifications and upgrades of the apparatus and experimental configurations, aiming to measure kaonic deuterium x-rays with comparable precision to the kaonic hydrogen measurement.
The key requirements were obtained with a GEANT4 Monte Carlo (MC) simulation: a new generation of large area X-ray detectors (SDD), with trigger capabilities to allow background suppression, with good energy (FWHM $\simeq$~150 eV at 6.4 keV) and timing resolutions (500~ns) and stable working conditions, even in a high accelerator background environment; dedicated veto detector systems to improve the signal-to-background ratio by at least one order of magnitude, compared to the kaonic hydrogen measurement performed by SIDDHARTA; a cryogenic light-weight gaseous target system to cool the gas down to 20 K, while the pressure could be tuned up to 2 bars to optimize the kaons stopping efficiency.

\section{The SIDDHARTA-2 setup}

\begin{figure}[ht]
\centering
\mbox{\includegraphics[width=14 cm]{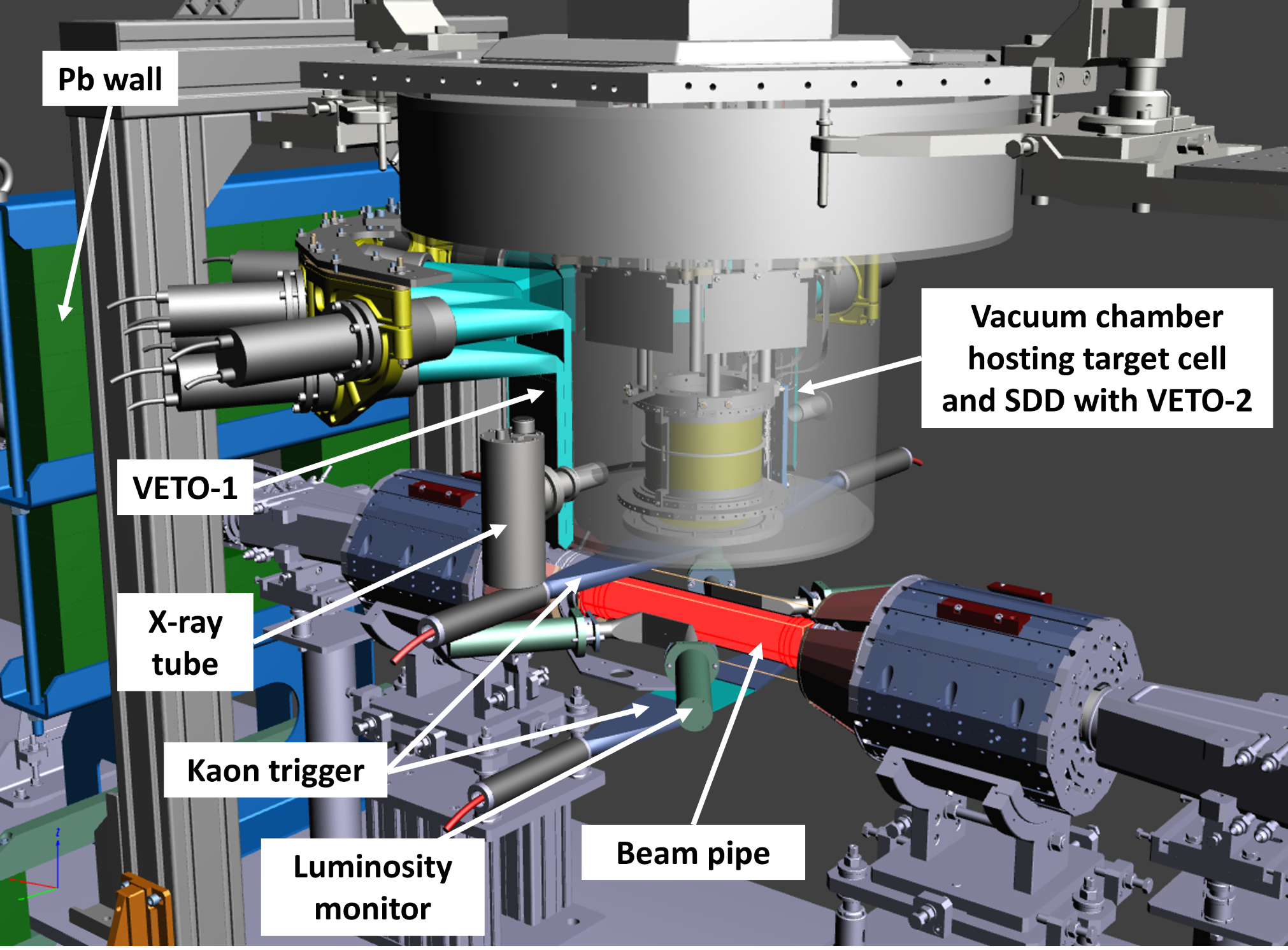}}
\caption{Schematic view of the SIDDHARTA-2 apparatus installed on the IP1 of the DA$\Phi$NE collider.}
\label{setup02}
\end{figure}

The SIDDHARTA-2 setup was mounted in the interaction point (IP1) of the DA$\Phi$NE collider in 2022. Low-momentum negatively charged kaons from $\phi$(1020) meson decay pass through the beam pipe, are degraded in energy down to a few MeV by a Mylar degrader and partially stopped in a gaseous target placed at a short distance above the beam pipe. The X-rays produced by de-excitation of the exotic atoms are detected by X-ray Silicon Drift Detectors (SDDs). A sketch of the apparatus is shown in Fig. \ref{setup02} where the main components are: beam pipe, kaon trigger, luminosity monitor, cylindrical vacuum chamber holding the cryogenic target and the X-ray detectors (SDDs), and veto systems. 

The vacuum chamber is evacuated below $\mathrm{10^{-6}\, mbar\,}$ by a turbo-molecular pump (TURBOVAC 350i/ix) with a pumping speed of 300~l/s and  oil-free isolated dry scroll pump (SCROLLVAC) from Leybold. A system composed of two X-ray tubes, type Jupiter 5000 Series at 50kV from Oxford Instruments, is employed for the in-situ calibration of the SDDs, taking advantage of the excitation of the fluorescence lines from high-purity titanium-copper (Ti-Cu) foils mounted on the target. A lead table and two lead walls complete the structures to shield the apparatus from particles, mostly Minimum Ionizing Particles (MIPs), lost from the $e^+e^-$ rings. The components of the apparatus are described in more detail in the following subsections.

\subsection{Beam pipe and Mylar degrader}\label{IP_section}

\begin{figure}[htbp]
\centering
\mbox{\includegraphics[width=10cm]{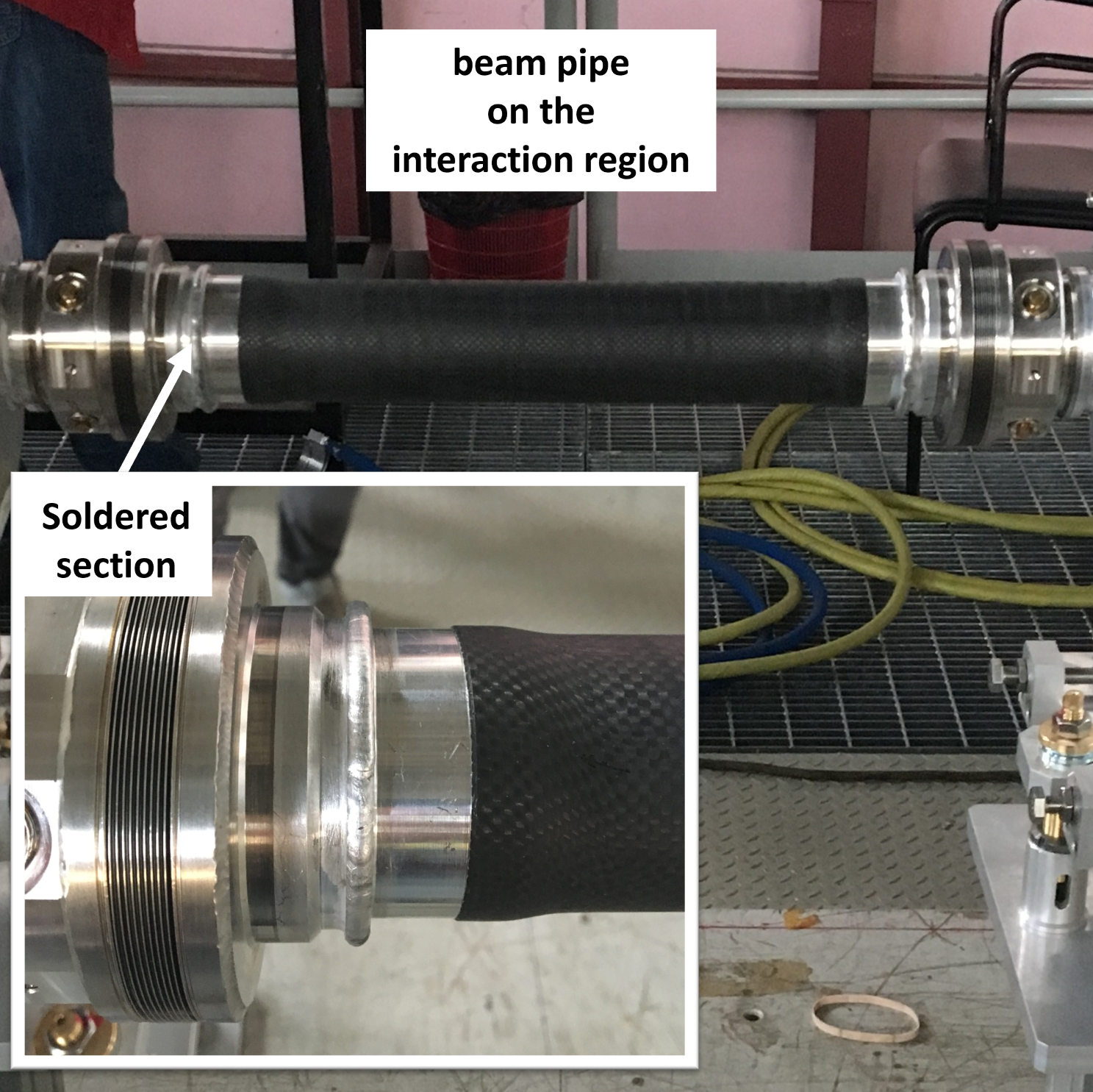}}
\caption{The interaction region (IP1) after installation of the beam pipe for the SIDDHARTA-2 experiment.}
\label{ir2_dafne}
\end{figure}

A cylindrical beam pipe made of 150 $\mathrm{\mu m}$ thick aluminum walls, with an internal diameter of 58~mm and a length of 380~mm, covered by an external layer of 500~$\mathrm{\mu m}$ carbon fiber, was designed and built (see Fig.\ref{ir2_dafne}). All possible discontinuities have been avoided in order to keep the ring coupling impedance low. The number of bellows has also been limited to the strictly necessary to compensate for thermal strain and mechanical misalignment. The connection of the beam pipe to the accelerator beam line in the interaction region was soldered in order to minimize the background (no high-Z materials near the IP1, e.g. vacuum flanges). 

\begin{figure}[htbp]
\centering
\mbox{\includegraphics[width=14 cm]{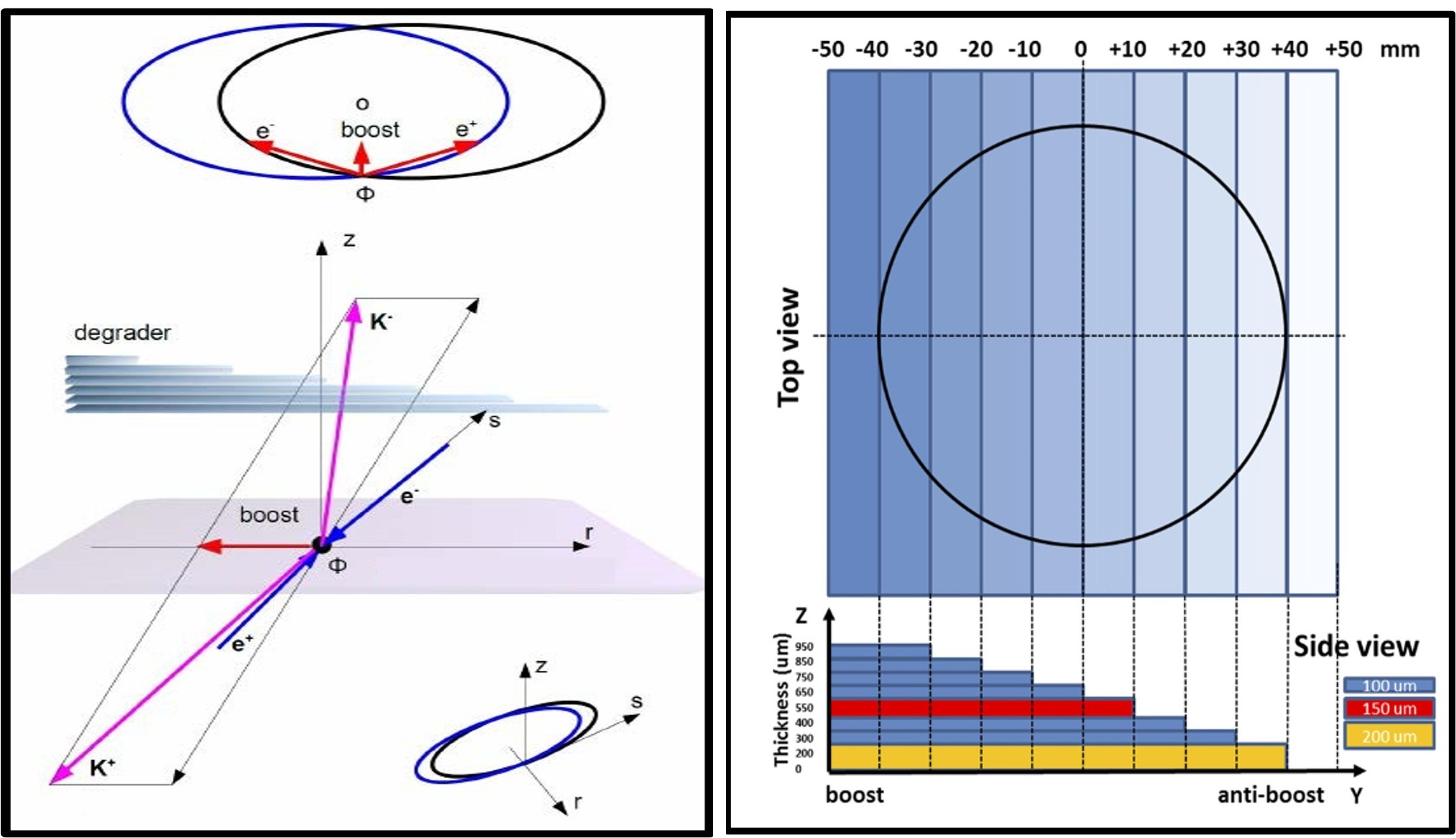}}
\caption{\small{Left: The electron-positron crossing beams in the IR1 indicating the boost direction to the center of the collider. Right: an example of the Mylar degrader configuration; the circle represents the size of the entrance window of the vacuum chamber; direction “Y” points to the exterior of the DA$\Phi$NE ring, corresponding to the anti-boost side for kaons. The degrader with variable thicknesses is shown in the lower part of the figure (from reference \cite{Sirghi:2022wbj}). }}.
\label{degrader}
\end{figure}

Since the electron and positron beams cross at the interaction point, with an angle of 50~mrad, the center of mass of the subsequent $K^+K^-$ system receives a boost towards the center of the collider, which is reflected in the momentum distribution of the kaons (see Fig.\ref{degrader}).\\ To compensate for this effect and to obtain a uniform stopping  distribution of the kaons inside the target, a step-wise Mylar degrader is used and placed below the upper scintillator of the kaon trigger. The degrader used during the helium measurements is schematically shown, as an example, in Fig. \ref{degrader}. It consists of 8 Mylar strips of $\mathrm{1\times9\,\, cm^2}$ each varying from $\mathrm{100\, \mu m}$ to $\mathrm{200\, \mu m}$ in thickness ($\mathrm{550\, \mu m}$ in the central position, namely Y=0 in Fig.~\ref{degrader}). The overall thickness of the eight-step degrader ranges from $\mathrm{200\,\mu m}$ to $\mathrm{950\, \mu m}$.\\

\begin{figure}[h]
\centering
\mbox{\includegraphics[width=14cm]{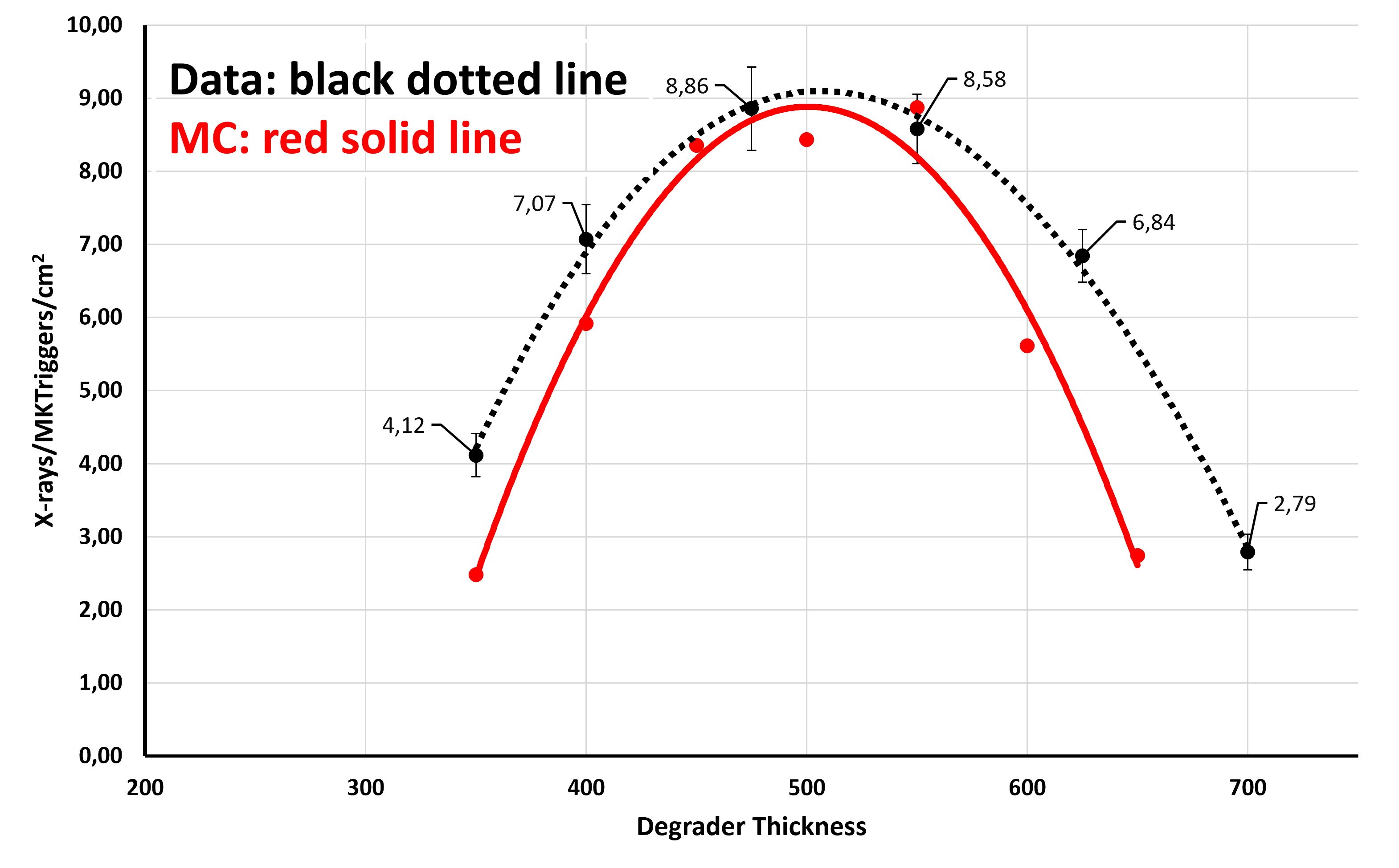}}
\caption{Degrader optimization curves: kaonic helium-4 ($K^4He$) $3d\rightarrow2p$ signal normalized to the numbers of the triggered kaons and effective detection surface (black dotted line) vs. degrader thickness in $\mu$m, together with the MC predictions (red solid line).}
\label{deg_curve}
\end{figure}

To optimize the shape and thickness of the degrader, GEANT4 Monte Carlo (MC) simulations have been performed, as well as an experimental fine tuning based on the intensities of kaonic helium X-rays. For different degrader configurations, the number of kaonic helium-4 ($K^4He$) $3d\rightarrow2p$ transition of  x-ray events, normalized to the numbers of the triggered kaons and effective detection surface was recorded. Its trend as a function of the degrader central thickness is shown in Fig.~\ref{deg_curve}  together with the MC predictions. The larger width of the results obtained from the measurements compared with the MC is explained by the uncertainty in the measurement of the overall material budget in the setup. The two distributions are in good agreement for the optimal degrader configuration, at the maximum of the curves.    

This optimization is an important and delicate operation since, as clearly shown by the curve, even a small variation of about 200 $\mu$m can drastically reduce the kaonic atom signal.

\subsection{Kaon trigger detector and luminosity monitor}\label{lumi_section}

\begin{figure}[ht]
\centering
\mbox{\includegraphics[width=14 cm]{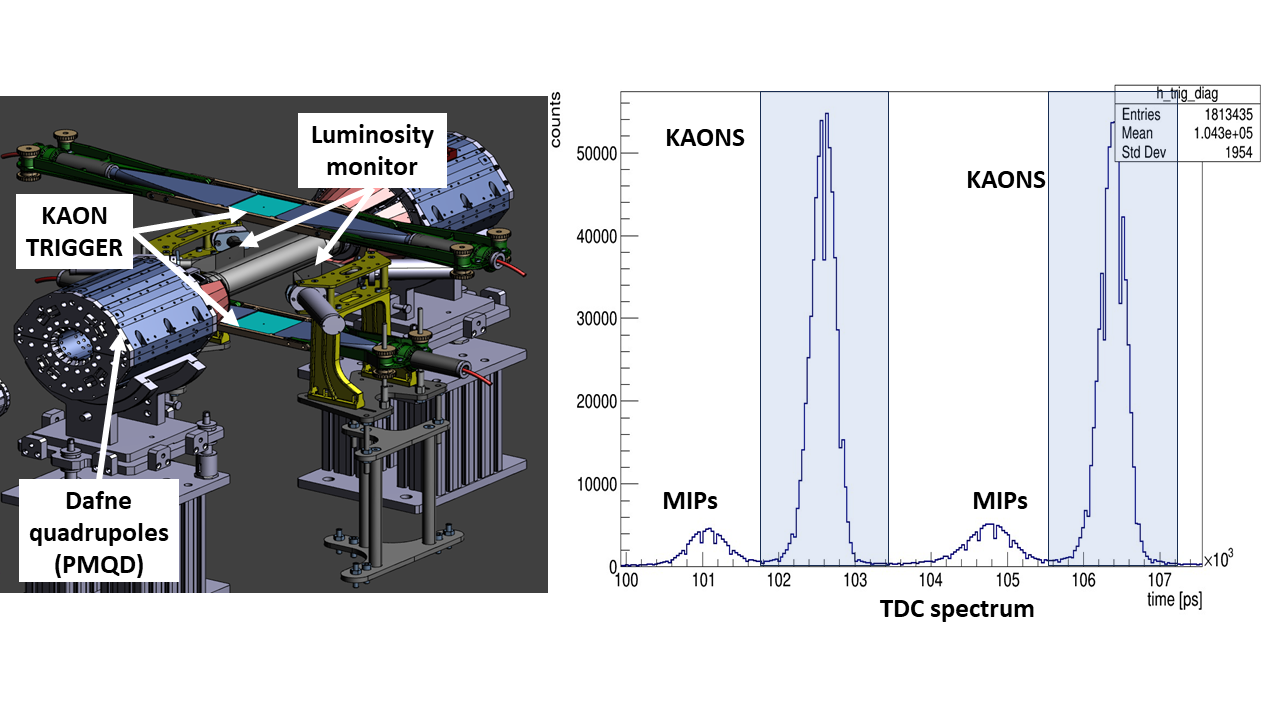}}
\caption{\small{Left: the kaon trigger placed in the vertical plane and the luminosity monitor on the horizontal plane of the interaction point. Right: TDC spectrum with selected time window used for kaon-identified events.}
\label{kt_lm}}
\end{figure}

The logic of the trigger system consists of the coincidence between the SDD signal and the signal associated with the production of a $K^+K^-$pair. The identification of charged kaon pairs from the $\phi$(1020) decay is based on the low momentum back-to-back kaons. The kaon pairs are detected as coincidence events with two plastic scintillators (EJ-200 from Scionix) with a size of 100$\times$100$\times$1.5 mm$^3$ read by Photo-Multiplier Tubes (PMTs) R4998 from Hamamatsu, mounted above and bellow the beam pipe, as shown in Fig.~\ref{kt_lm}. Only X-ray SDD signals falling in a 5 $\mu$s time window in coincidence with a trigger signal are selected, rejecting a substantial portion of background. The time window width was tuned to enable the front-end electronics to process and acquire the signals.
Background events, also counted by the kaon detector, are charged particles from $\phi$(1020) decay, (e.g. to $\pi^+$, $\pi^-$, $\pi^0$) as well as from the decay of neutral particles and electromagnetic showers from the 510~MeV electrons and positrons circulating in the main ring, hitting the beam pipe. These are fast MIPs, while the charged kaons are slow and highly ionizing. The charged kaons arrive about 1.5~ns later than the MIPs and deposit ten times more energy in the scintillators. 

The time difference between the signals on each scintillator and the DA$\Phi$NE radio-frequency (RF$\simeq$~368~MHz) signal is used to discriminate, via time of flight (ToF), whether an event on the kaon detector is related to a kaon or to a MIP entering the target cell. The double kaon/MIPs structure in the plot of the TDC spectrum is due to the usage of the RF/2 signal as a time reference, processed by a constant fraction discriminators (CFD) with a maximum bandwidth of 200 MHz. From Fig. \ref{kt_lm}, one can see the kaon-related events, corresponding to the two main peaks, which are easily discriminated from the MIPs-related ones. Among all triggered events, only the kaon-identified events, which are contained in the time windows within the dark areas in Fig. \ref{kt_lm}, are finally accepted. The kaon trigger is used to achieve a background reduction factor about of 10$^5$, defined as the ratio between the number of events after the trigger request and that of the raw spectrum, without which signals from kaonic atoms would be impossible to observe.\\
The luminosity monitor, based on the technology developed by J-PET for imaging of electron–positron annihilation~\cite{jpet1,jpet2,jpet3}, is employed to evaluate both the machine luminosity and background level via time of flight (ToF) measurements of kaons and MIPs in the horizontal plane of the interaction point (IP1) (see Fig. \ref{kt_lm}). It consists of two parallel detection modules located at the left and right sides of the beam pipe, in the collision plane, corresponding to the $\phi$ boost and anti-boost directions, at a distance of 7.2 cm from the IP1. Each module consists of a 80$\times$40$\times$2 mm$^3$ BC-408 plastic scintillator, from Saint-Gobain Crystals, coupled at both ends to fast R4998 Hamamatsu PMTs through 6~cm fish-tail plastic light guides. The light guides scintillator contact surfaces have an angle of 38 degrees to fulfill the geometrical constraints of the apparatus \cite{Skurzok:2020phi}.

\begin{figure}[htbp]
\centering
\mbox{\includegraphics[width=12 cm]{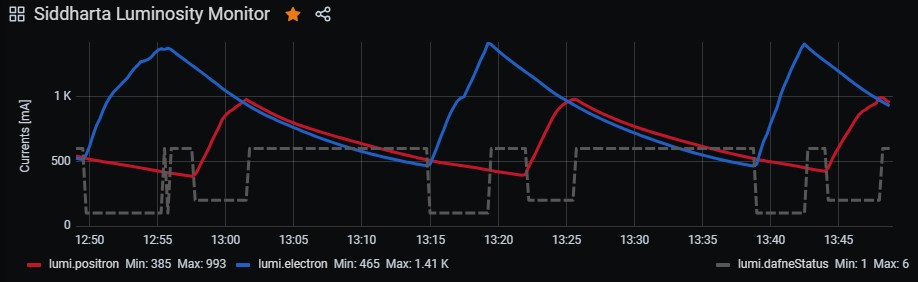}}
\mbox{\includegraphics[width=12 cm]{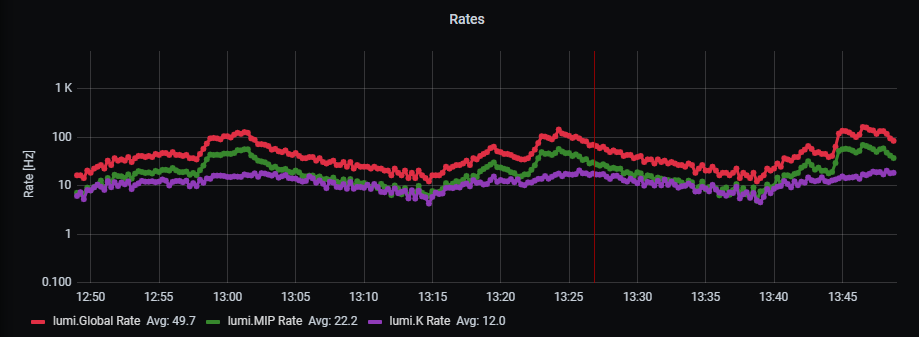}}
\caption{One hour record of the beam parameters during the kaonic helium data taking. Top: red line - positron and blue line -electron beams, the status of the accelerator - dashed line; Bottom: overall coincidence rates - red line, counting rates for online selected kaons - green line and MIPs - magenta line from the luminosity monitor.} 
\label{slow1}
\end{figure}

The luminosity measurement is one of the most important diagnostic tools to optimize the performances of the collider.  A data acquisition system (DAQ) measure the instantaneous luminosity based on the number of kaon coincidences in a given time interval, normalized by a scale factor evaluated by Monte Carlo simulations. The time interval has been fixed to 15s to ensure accuracy of a few percent on the measurement when the luminosity is ranging between 0.5$\times$10$^{31}$ cm$^{-2}$s$^{-1}$ and 20$\times$10$^{31}$ cm$^{-2}$s$^{-1}$.
Combined with the fast information from the SDD detectors the ratio kaons/MIPs from this real-time data is used for machine fine-tuning during operations, to check the machine setup variation, and to record the long-term performances. An application has been developed under the Slow Control System to drive data acquisition. A screenshot of the graphical user interface is presented in Fig.~\ref{slow1}, whose central panel shows the diagnostic behavior of the colliding beams. 
\begin{figure}[htbp]
\centering
\mbox{\includegraphics[width=12 cm]{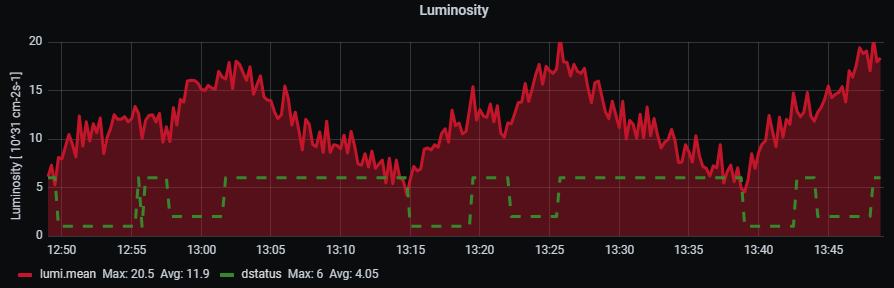}}
\caption{The estimated instantaneous luminosity, in units of 10$^{31}$ cm$^{-2}$s$^{-1}$, for one hour of the kaonic helium data on 6 May 2023. The status of the accelerator in the same period - dashed green line.} 
\label{slow2}
\end{figure}
The application  track beam currents, coincidence rates between the two scintillators, counting rates for online selected kaons and MIPs and estimated instantaneous luminosity, as shown in Fig.~\ref{slow2}.

\subsection{Cryogenic target cell}\label{target_section}

\begin{figure}[htbp]
\centering
\mbox{\includegraphics[width=12 cm]{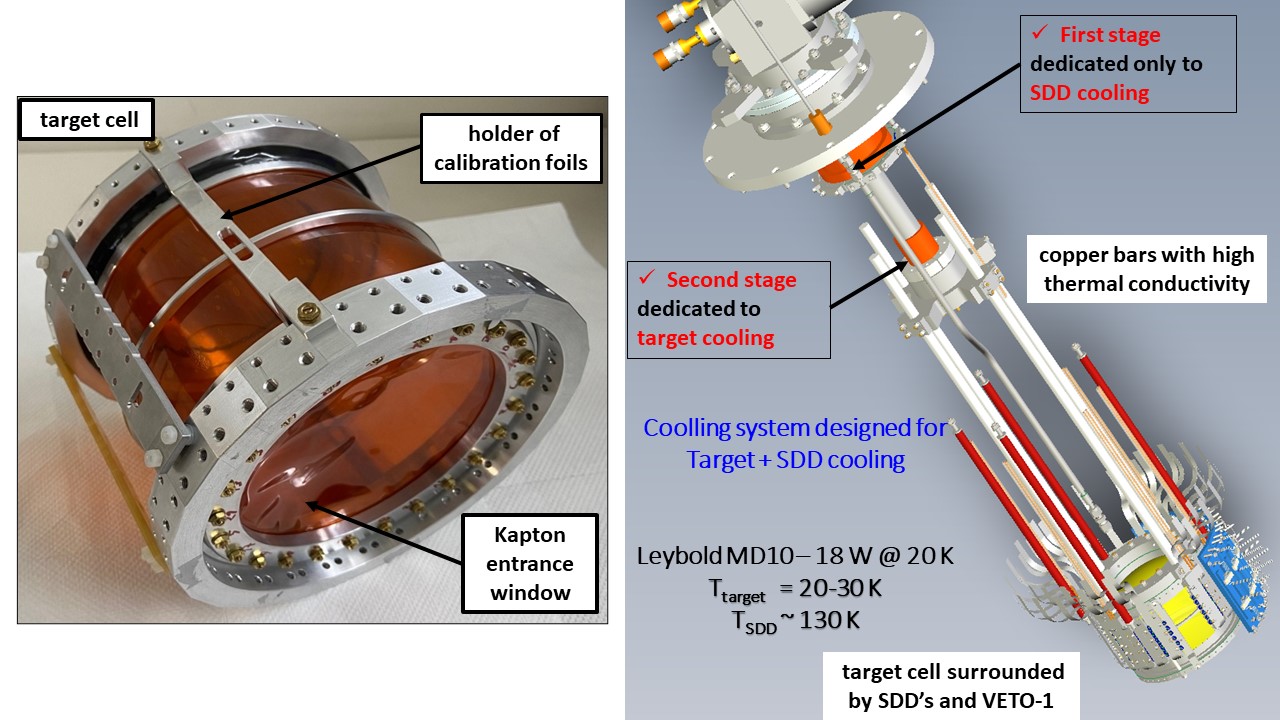}}
\caption{Left: the SIDDHARTA-2 target cell made of Kapton walls. Right: the cooling system for the SDD arrays and target cell.}
\label{target01}
\end{figure}

The cryogenic target cell is a cylindrical volume with a diameter of 144~mm and a height of 125~mm. The side walls are made of two layers of 75~$\mathrm{\mu m}$ Kapton, (C$_{22}$H$_{10}$N$_2$O$_5$) that can remain stable for a wide range of temperatures, with a reinforcement structure made of high purity aluminum, as shown in Fig. \ref{target01}. It has a 125~$\mathrm{\mu m}$ Kapton entrance window for the kaons and a 100~$\mathrm{\mu m}$ thick titanium (Ti) top layer, which is useful for calibration purposes, together with additional high purity titanium-copper (Ti-Cu) strips placed on holders on the target cell walls. The cooling and mounting structure of the SDDs are used to reinforce the target cell in the longitudinal direction, as shown in Fig.~\ref{target01}. \\
The presence of the strong electric field causes Stark mixing effects which describes the shifting or splitting of atomic states for an atom in an external electric field. To achieve the balance between high kaon stopping in the target and the decreasing of the X-ray yield due to the Stark mixing, the target density should be tuned to compensate for the two effects \cite{koike1996}.
A closed-cycle helium refrigeration system is used to keep the target cell between 20–30~K with maximum pressure of 2~bar. The cryogenic refrigeration system comprises the compressor unit (COOLPAK 6000-1) and the cold head (model COOLPOWER 10 MD) connected with flex-lines. A compressor unit compresses the helium gas which then expands in the cold head to produce low temperatures. The two-stage cold head with a flange of 190 mm diameter provides enough cooling power for the SDD detectors from the first stage (115 W at 80 K) and for the target from the second stage (18 W at 20 K). A heating system was also installed on the target, allowing for operating the target cell at different temperatures and pressures, depending on the gas mixtures.\\

\subsection{The SDD detector system}\label{SDD_section}

\begin{figure}[ht]
\centering
\mbox{\includegraphics[width=12 cm]{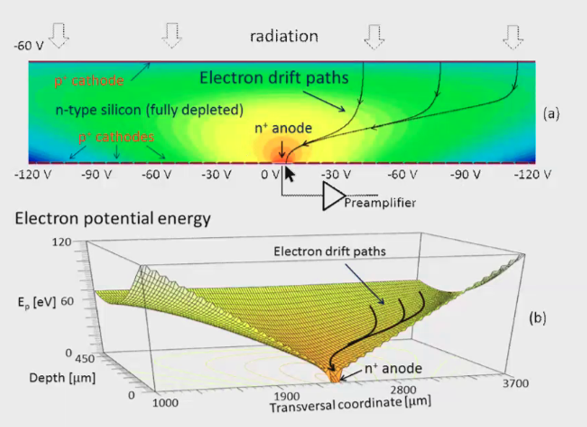}}
\caption{The drift field is determined by the potential applied to the voltage divider for an SDD with uniform entrance window. Charge collection and the electron drift paths in the anode region.}
\label{sdd1}
\end{figure}

Silicon Drift Detectors (SDDs) were first proposed by E. Gatti and P. Rehak in 1984 \cite{Gatti:1984}. The basic idea of the side-ward depletion – the revolutionary concept at the basis of the SDD and of the related devices – is to fully deplete the semiconductor substrate through a point-like "virtual contact", as it is called in the original paper. The key feature of SDDs, with respect to a conventional PIN diode of equivalent active area and thickness, is its extremely low readout capacitance (100 fF). This capacitance is determined by the small n+ anode area and is essentially independent on the detector size, a crucial improvement as compared to a simple PIN photo diode. Therefore an ultimate energy resolution can be in principle achieved also for large area detectors. A further improvement is the circular shape, in which the SDD has a small central anode with a set of concentric p+ rings on the bottom side and a large continuous p+ electrode (called “continuous back-plane”) on the upper side. Since the detectors for spectroscopy are intended to measure only the energy and not the interaction point of the incident radiation, the drift field is designed to have azimuthal symmetry thanks to circular concentric drift rings. The potential gradient applied on the upper side sets the drift field for electrons, and the potential on the continuous back-plane guarantees the full depletion also in the anode region. The electrons produced anywhere within the detector active area are collected at the central annular anode (see Fig.~\ref{sdd1}). The back plane is realized with a uniform p+ implant, which acts as a thin entrance window for the radiation.

\begin{figure}[ht]
\centering
\mbox{\includegraphics[width=12 cm]{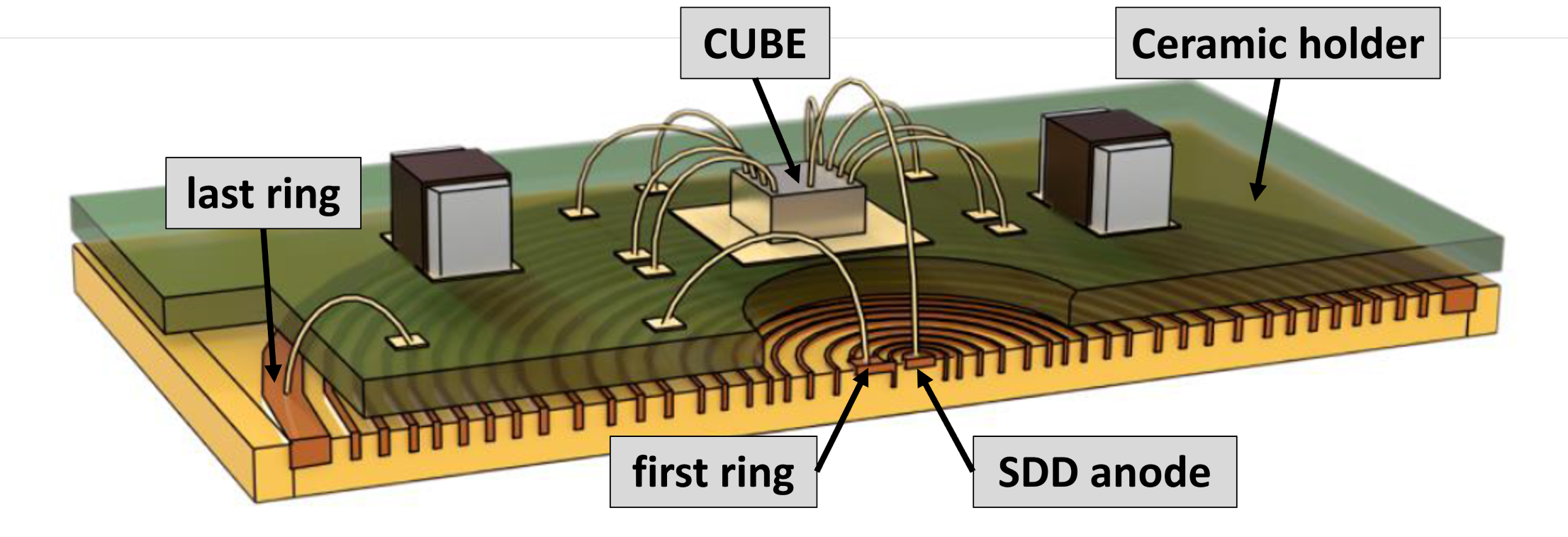}}
\caption{ Schematic layout of a cylindrical Silicon Drift Detector. The CUBE pre-amplifier is installed close to the SDD anode. The ceramic holder board provides cooling contact and voltage supplies for the circular drift rings.}
\label{sddcube}
\end{figure}
\begin{figure}[ht]
\centering
\mbox{\includegraphics[width=14 cm]{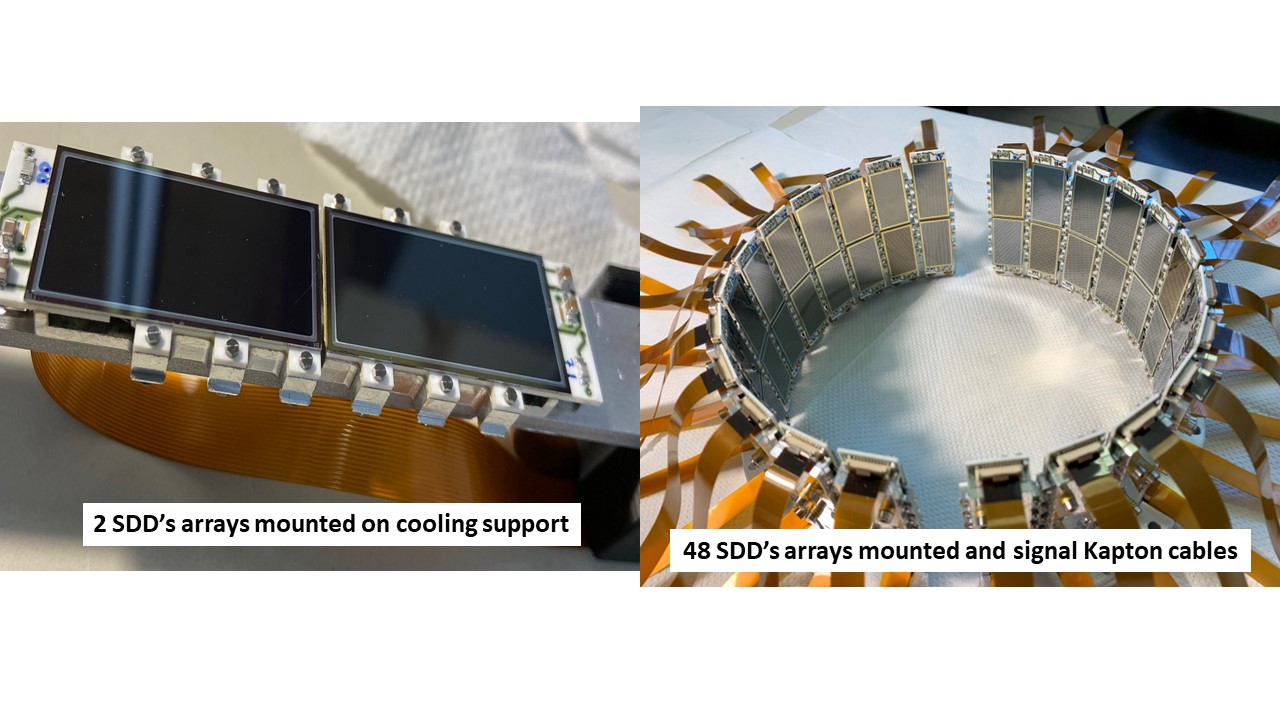}}
\caption{Left: the two SDD arrays mounted on the cooling holder. Right: all 48 SDD arrays ready to be mounted around the target.}
\label{sdd01}
\end{figure}

The large area monolithic SDD arrays have been developed by Fondazione Bruno Kessler (FBK), for the SIDDHARTA-2 experiment. The silicon wafer is glued on an alumina board which provides the polarization to the units via an external voltage. The charges generated by the X-ray absorption within the silicon bulk are collected by a point-like central anode and amplified through a closely bonded CMOS low-noise, charge-sensitive pre-amplifier (CUBE) \cite{cube:2010}. The CUBE pre-amplifier, installed close to the SDD anode, has been designed to render the device performance independent from the applied bias voltages and to increase its stability, even when exposed to high charged particle rates generated by the collider operations (see Fig.~\ref{sddcube}). This solution allows to operate the device at lower cryogenic temperature than the external J-FET pre-amplification stage used previously in SIDDHARTA, thus obtaining systematically lower electronic noise. 
Each 450~$\mathrm{\mu m}$ thick silicon array consists of 8 single cells arranged in a 2$\times$4 matrix, with a total size of  34$\times$18~$\mathrm{mm^2}$ (including dead areas). The design of the array and of the carrier has been optimized in order to maximize the total active area of 5.12~$\mathrm{cm^2}$ as shown in  Fig.~\ref{sdd01}. 

The ceramic board provides cooling contact, and voltage supplies for all 8 SDD cells. The output of the CUBE pre-amplifiers is connected to a common ASIC, called SFERA (SDDs Front-End Readout ASIC), a 16-channel integrated circuit performing analog shaping and peak detection of the signals~\cite{Quaglia:2016uox, Schembari:2016IEE}. The SFERA main shaper is characterized by a $9^{th}$ order semi-Gaussian complex-conjugate poles filter with discriminating peaking times (from 500 ns to 6 $\mu$s) and gains, while the fast shaper has a fixed 200 ns peaking time and is used for pile-up rejection. Each ASIC handles the signals produced by 16 units and provides the DAQ chain with the individual amplitude and timing information. The spectroscopic performances of this system have been optimized in our laboratory \cite{Miliucci:2019mdpi} and tested during the beam commissioning phase of DA$\Phi$NE, started in early 2020, with very good energy resolution of $\mathrm{157.8\pm0.3^ {+0.2}_{-0.2}\,eV}$ at $\mathrm{6.4\,keV}$ and linearity at the level of 2-3~eV \cite{Miliucci:2021wbj}, proving to be a highly efficient detector system.\\

\subsection{Veto Detector Systems}\label{veto_section}

\begin{figure}[h]
\centering
\mbox{\includegraphics[width=12 cm]{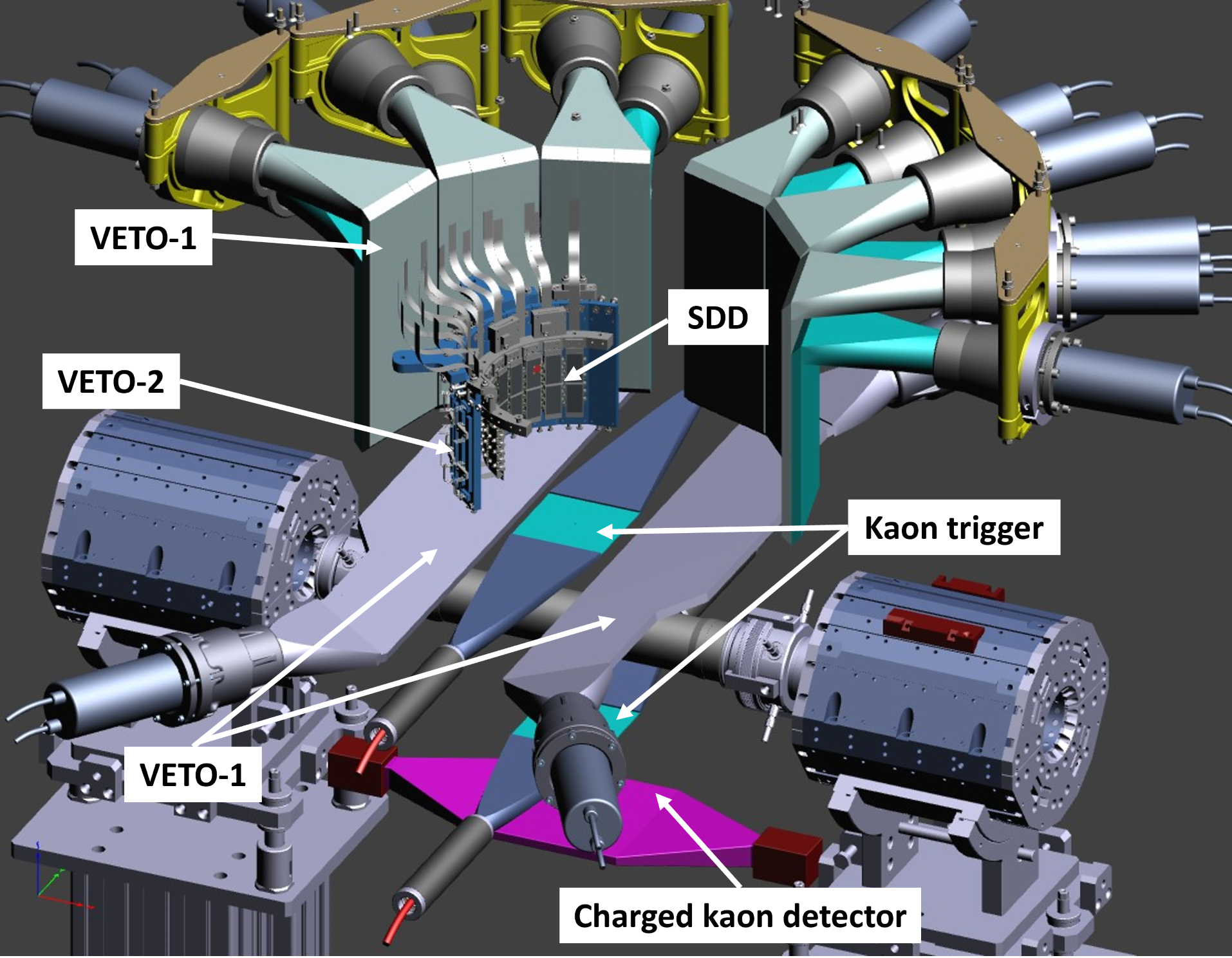}}
\caption{\small{Schematic view of the veto systems installed in SIDDHARTA-2. The Veto-1, an outer barrel of scintillators and two additional scintillators placed under the vacuum chamber at the sides of the kaon trigger. The Veto-2, an inner ring of plastic scintillation tiles, is placed at a short distance behind the SDD's. The bottom kaon detector (magenta) is located under the lower kaon trigger scintillator.}
\label{vetoall}}
\end{figure}

To measure kaonic deuterium transitions while dealing with low signal and significant background, it is crucial to select events that occur exclusively within the gas target and eliminate the hadronic background due to kaons stopping in the setup. Veto systems for SIDDHARTA-2 were designed, built and installed on the apparatus. They consist of an outer barrel of scintillator counters read out by PMTs (Veto-1), placed outside the vacuum chamber, and an inner ring of plastic scintillation tiles, read out by Silicon Photo-multipliers (SiPMs), placed as close as possible behind the SDDs (Veto-2), for charged particle tracking, as shown in Fig. \ref{vetoall}. To close the solid angle below the vacuum chamber an additional pair of plastic scintillators was installed.
Another system called bottom kaon detector, placed below the lower scintillator of the kaon trigger, distinguishes between events with either a $K^-$ or $K^+$ reaching the target cell via timing information. \\

\subsubsection{The Veto-1 system}\label{veto1_section}

\begin{figure}[h]
\centering
\mbox{\includegraphics[width=10 cm]{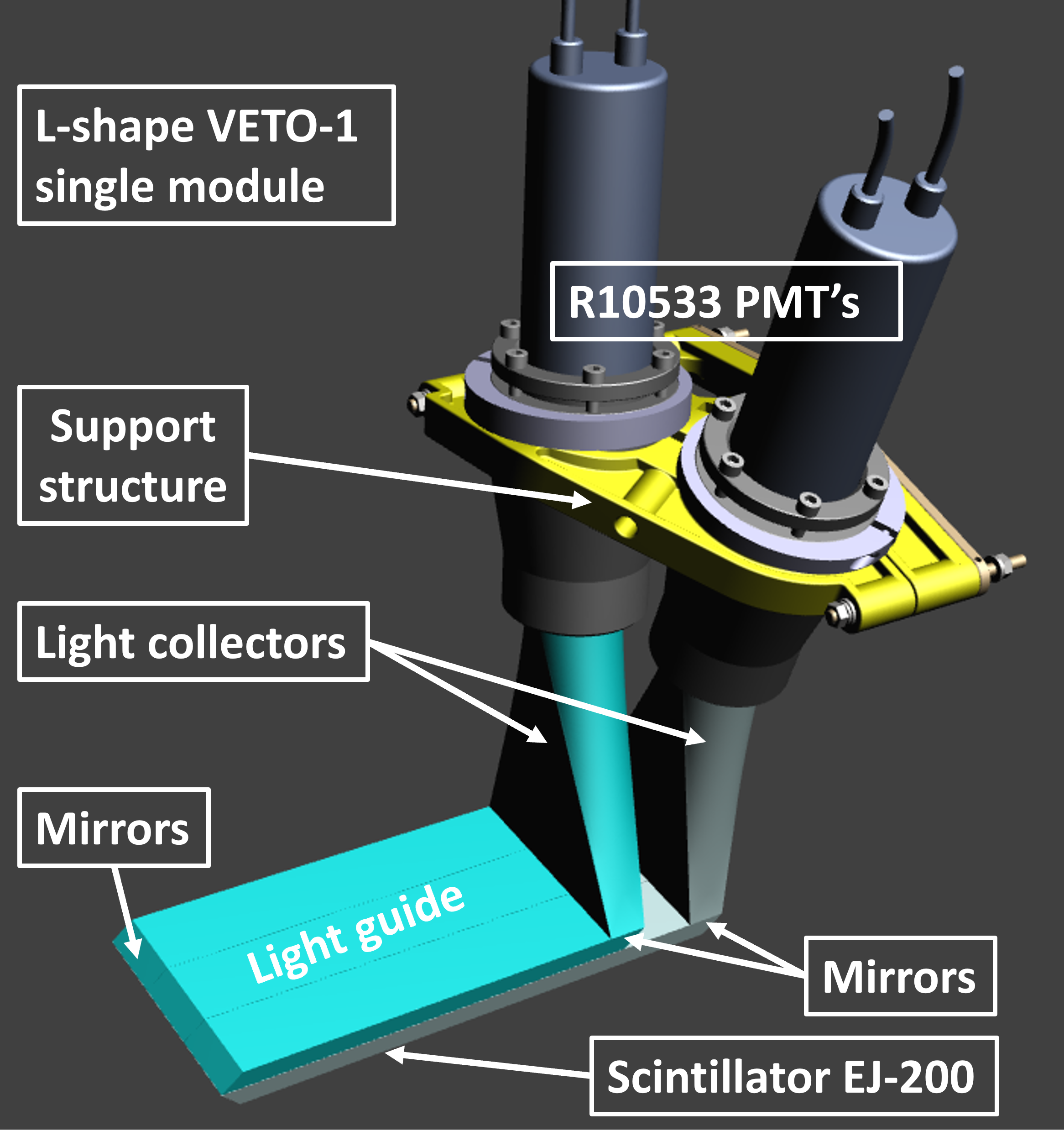}}
\caption{\small{The Veto-1 layout for single module. The L-shape design is required due to very small available space, involving mirrors to connect the scintillator and the light guide.}
\label{veto1_module}}
\end{figure}

The goal of the Veto-1 system is to reject the synchronous background produced by kaons stopping in the solid elements of the setup, mainly in the the target entrance window or in the vacuum chamber.
A $K^-$ stopping in the target gas, is captured by the gas atoms and undergoes an atomic cascade (or decays), with radiative and non-radiative transitions, until it is absorbed by the deuteron. Most of the final state channels include a charged pion. The pion has enough energy to pass through the SDD cooling support, the electronics, and the vacuum chamber, reaching the outer scintillator barrel of the Veto-1 system outside of the vacuum chamber. The same is valid for pions generated by kaons absorbed elsewhere in the setup. 
Based on the relatively long time that a kaon takes to stop in the target gas before it gets absorbed (compared to the short time e.g. the solid entrance window of the target or the vacuum chamber) one can build a veto counter by using this time information. 
The system takes advantage of the high (around 90$\%$) probability for the production of charged pions, following kaon absorption by the atomic nucleus. Given that pions coming from hadronic interaction easily pass through SDDs and vacuum chamber, the Veto-1 system consists of a barrel-shaped plastic scintillator, twelve modules in total surrounding the cylindrical vacuum chamber, each whit dimensions of 260$\times$110$\times$10 mm$^3$. Each plastic scintillator is read out by two Hamamatsu R10533 PMTs. The double read-out is necessary because, in addition to a better (mean time) resolution, the detector could provide information on the hit position (with an average uncertainty of 2~–~3~cm), by measuring the time difference between the two PMTs. 
Since in SIDDHARTA-2, the scintillators can be accessed only from one side, at a narrow transverse angle, a complex system of mirrors and light guides (Fig. \ref{veto1_module}) ensures the photon extraction, in order to fit the available space around the vacuum chamber, limited by the external lead shielding (\cite{Bazzi2013}).

\begin{figure}[ht]
\centering
\mbox{\includegraphics[width=12 cm]{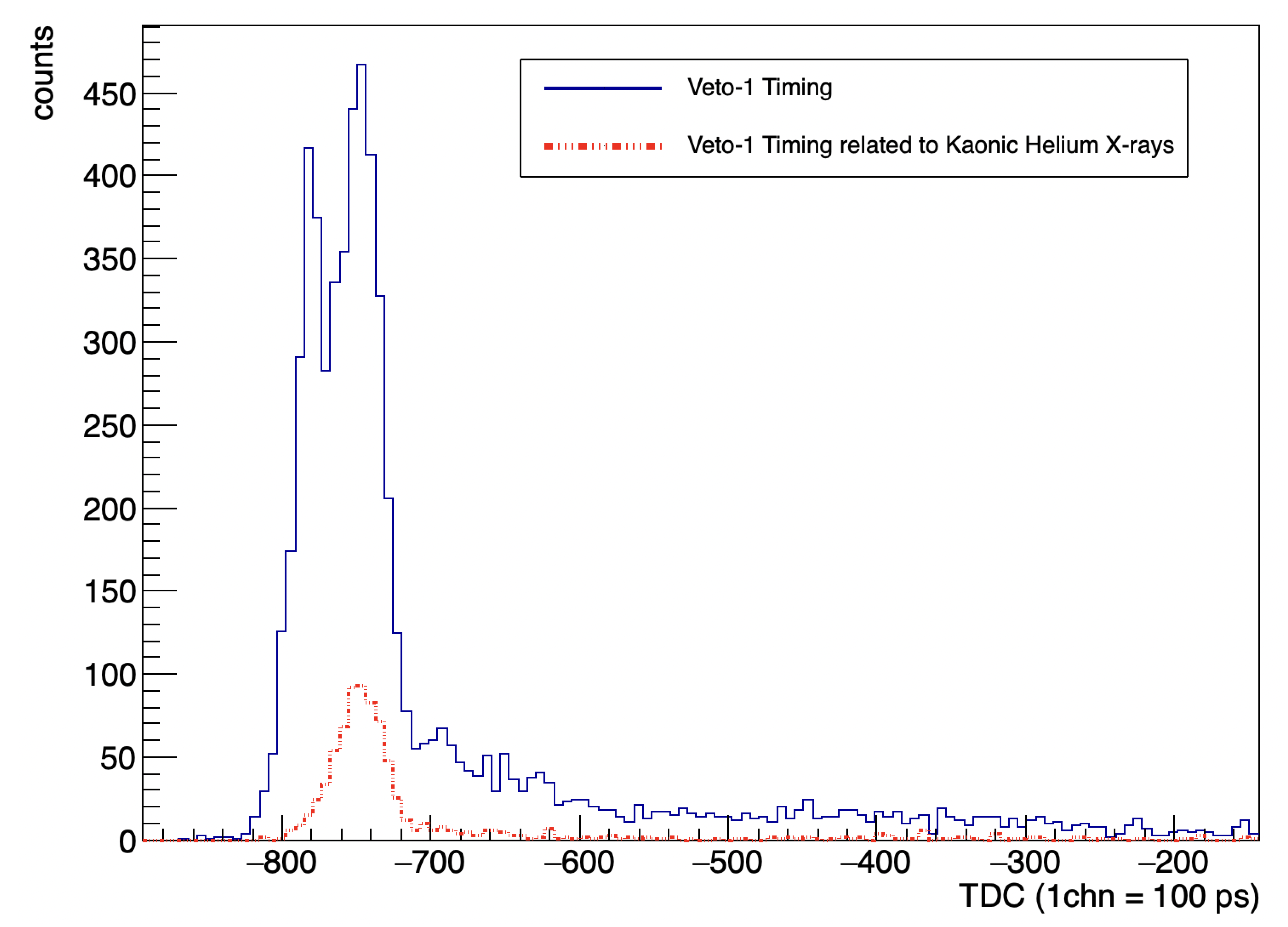}}
\caption{\small{Veto-1 time distribution from Time to Digital Converter (TDC), measured during kaonic helium run. The red dotted line represents the time distribution of events corresponding to kaonic helium atoms.}
\label{veto1_timing}}
\end{figure}

The Veto-1 system was characterized  at Paul Scherrer Institute (PSI) using a 170 MeV/c pion beam \cite{Bazzi2013}. The measured mean time resolution was (746$\pm$53) ps FWHM, while the efficiency was (96 $\pm$2)$\%$, in agreement with our requirements. In 2022 the full Veto-1 system was installed around the SIDDHARTA-2 vacuum chamber and was tested with the rest of the apparatus during the commissioning run of the experiment, when the  target cell was filled with helium. The operation of the full setup was verified with the high X-ray yield of kaonic helium.\\ 
The time difference between a pion induced signal in the Veto-1 scintillators and the kaon trigger signal is related to the moderation time of kaons in the gaseous target, allowing to distinguish events corresponding to kaons stopped in the gas from background events caused by the decay or absorption of kaons in the setup materials. This discrimination can be clearly seen in Figure \ref{veto1_timing}. The plot shows the time distribution, from the Time to Digital Converter (TDC), obtained during the kaonic helium run. The first peak corresponds to particles produced by $K^-$ absorption in the apparatus, such as the target entrance window, the degrader or the support frames. The second peak corresponds to $K^-$ stopped inside the gas target. The right tail is due to $K^+$ decay. The Veto-1 ability to provide timing information is crucial to improve the signal to background ratio, by rejecting events from $K^+$ decay or $K^-$ stopping outside the gas target.\\

\subsubsection{The Veto-2 system}\label{veto2_section}
\begin{figure}[ht]
\centering
\mbox{\includegraphics[width=6 cm]{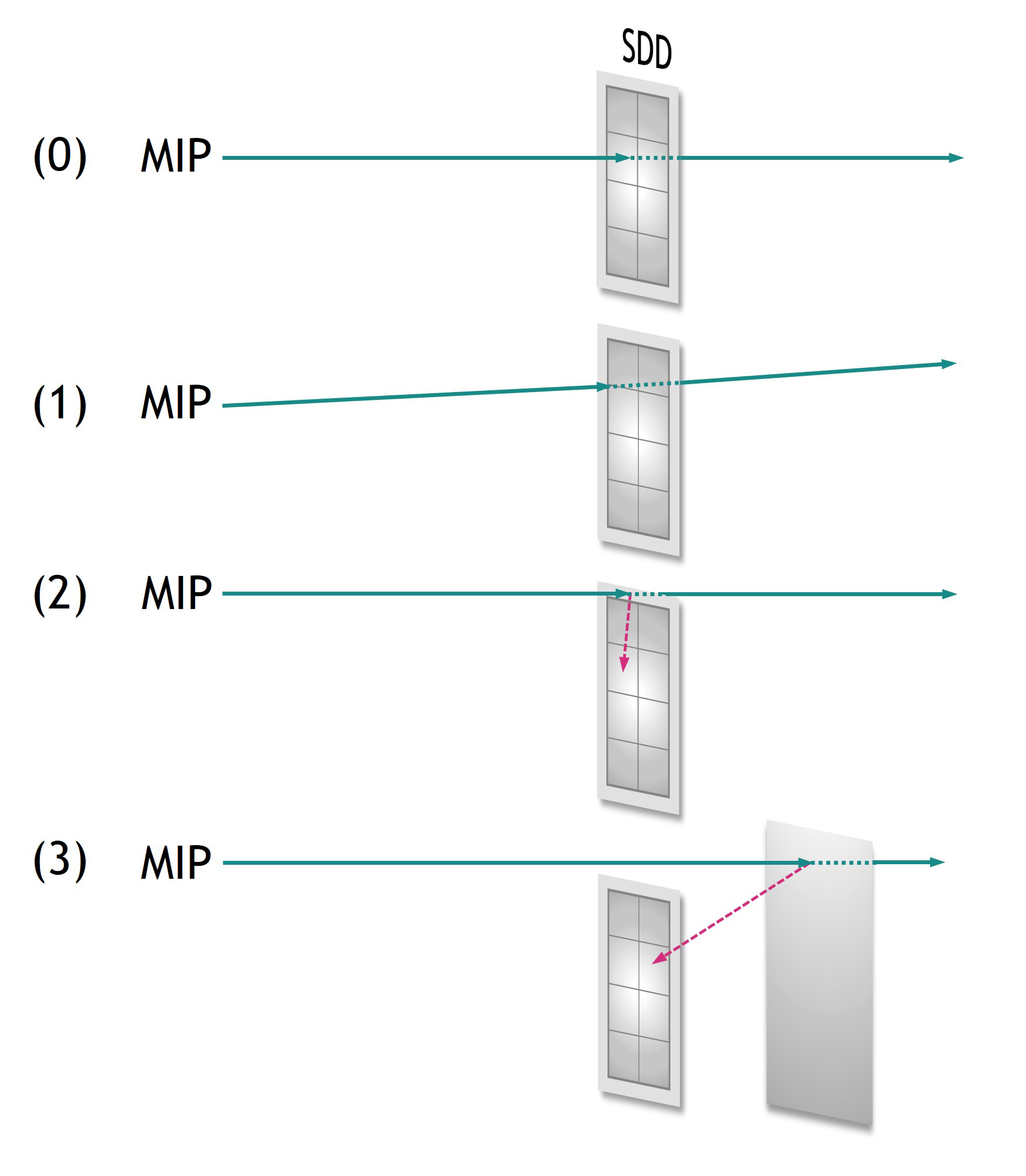}}
\caption{\small{Motivation for the Veto-2 system: (Case 0) The MIP passes through the SDD's active center, producing a large signal outside the region of interest (ROI). (1) The MIP hits the SDD on the edge of its active area, and a signal in the ROI can be produced. (2) Delta-rays hit the SDD. (3) Secondary X-rays or back-scattered electrons from the apparatus pass through the SDD.}
\label{veto2_sdd}}
\end{figure}

The kaons stopping inside the gaseous target produce, together with the kaonic atoms X-rays, several charged particles, mostly pions, protons as well as neutrons and gamma rays, following nuclear absorption. The goal of the Veto-2 system is to suppress this synchronous background. 
The charged particles can deliver signals in the SDDs in the same energy range as the relevant X-rays, thus representing an important background source to be identified and eliminated. The signal created by the charged particles depends on their path through the SDDs, as shown in Figure \ref{veto2_sdd} (\cite{Marlene:2023}).
A MIP passing the SDD on the edge of its active area deposits less energy than when passing through its center, which is also true for back-scattered electrons or secondary X-rays traversing the X-ray detectors. By using the spatial correlation between signals in the SDDs and the corresponding Veto-2 scintillators, this background source can be eliminated.

\begin{figure}[h]
\centering 
\mbox{\includegraphics[width=14 cm]{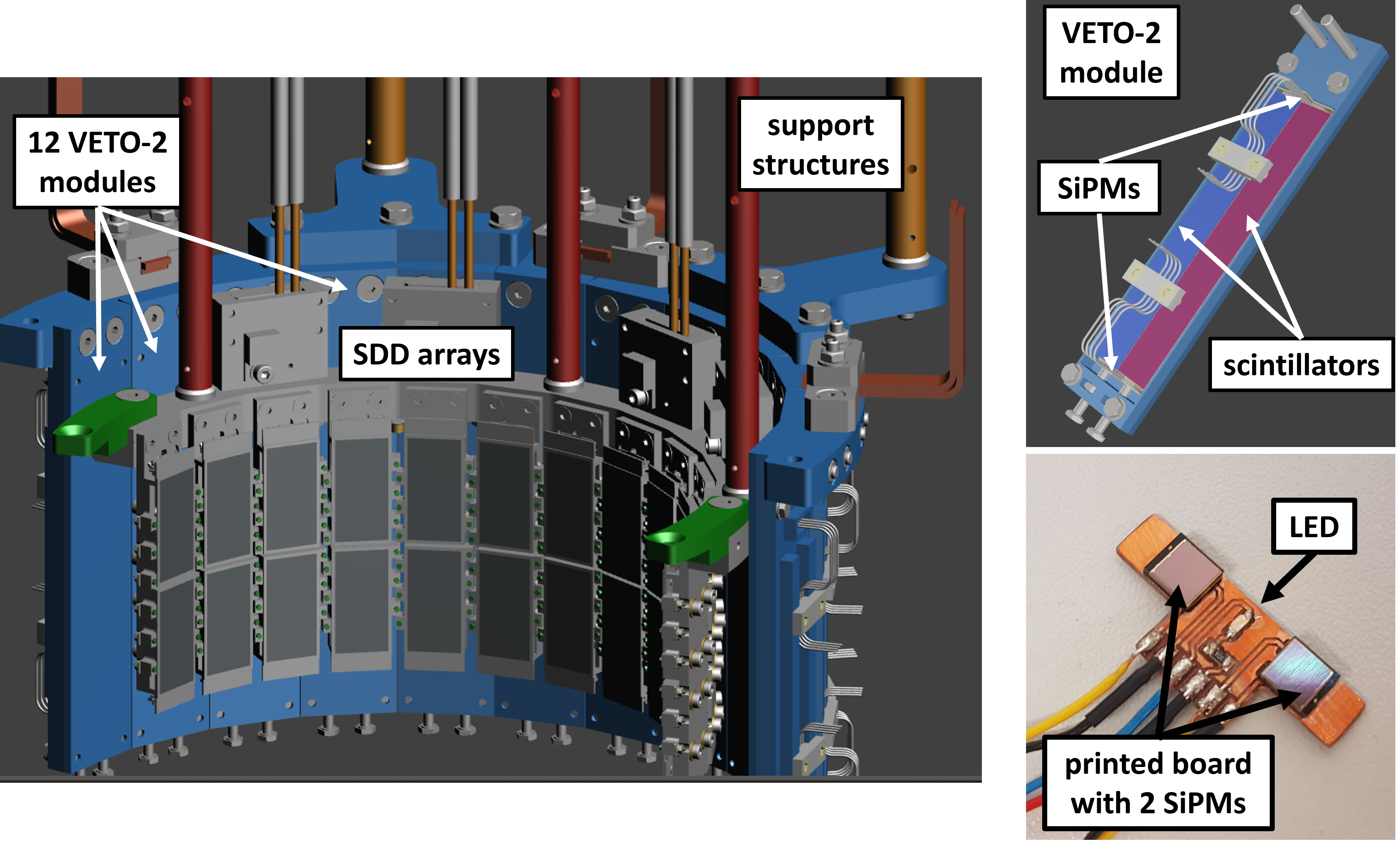}}
\caption{\small{The Veto-2 system designed for SIDDHARTA-2: half of the total Veto-2 units placed behind the SDDs (left side). Right side: single unit with two scintillators and the corresponding PCB board with two 4$\times$4 mm SiPMs; a blue LED is mounted in-between them for calibration of the system.}
\label{veto2}}
\end{figure}

The Veto-2 system consists of 24 units of 4 detectors each, resulting in 96 read-out channels. One detector module includes a 50$\times$12$\times$5 mm$^3$ plastic scintillator and a SiPM of size 4$\times$4 mm, manufactured by AdvanSiD. Optical grease and nylon screws couple the SiPMs to the scintillators \cite{Marlene;2018}. For calibration of the detectors, a blue LED is mounted in-between two SiPMs on one print board, as shown in Fig.~\ref{veto2}.

\begin{figure}[h]
\centering 
\mbox{\includegraphics[width=4 cm]{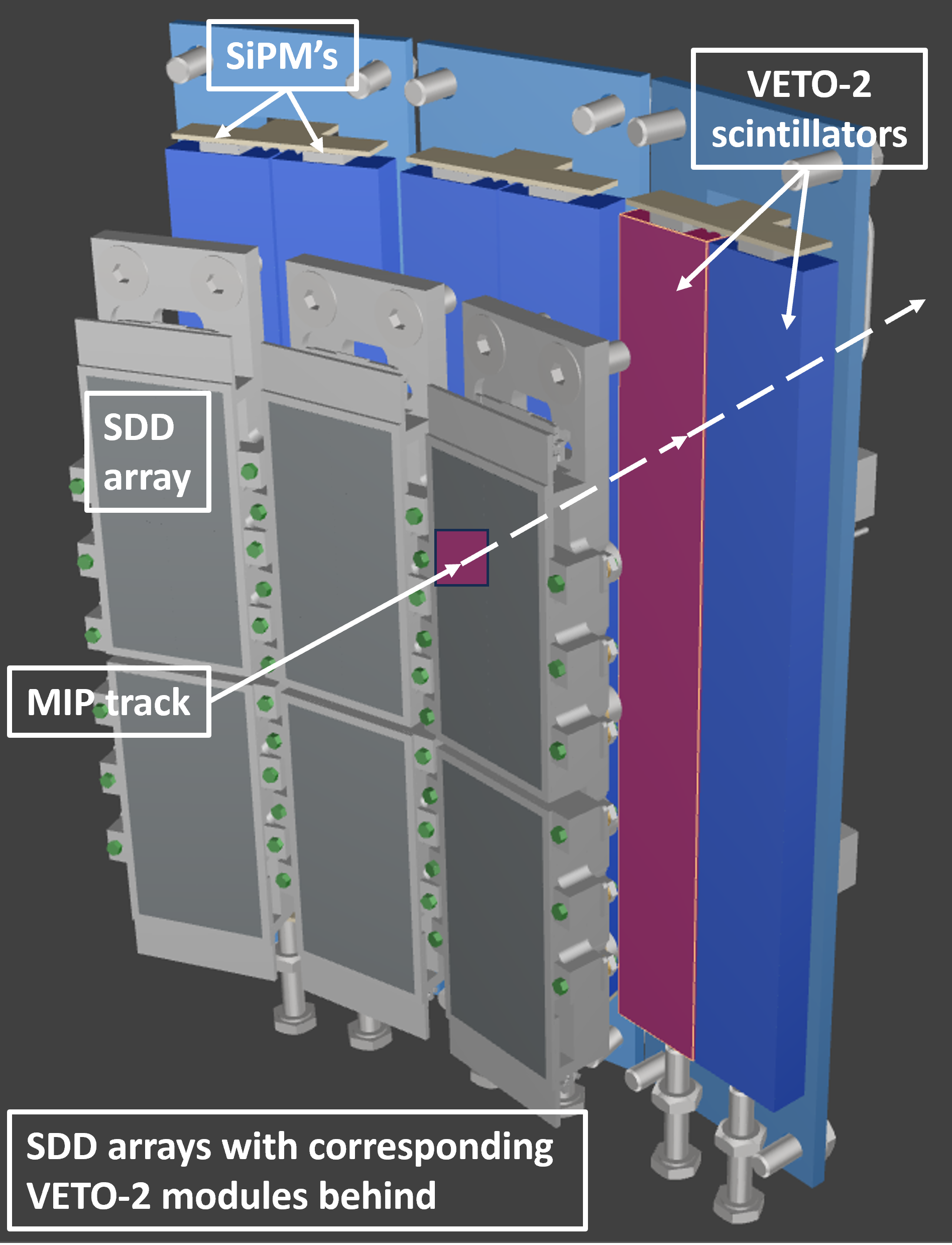}}
\mbox{\includegraphics[width=8 cm]{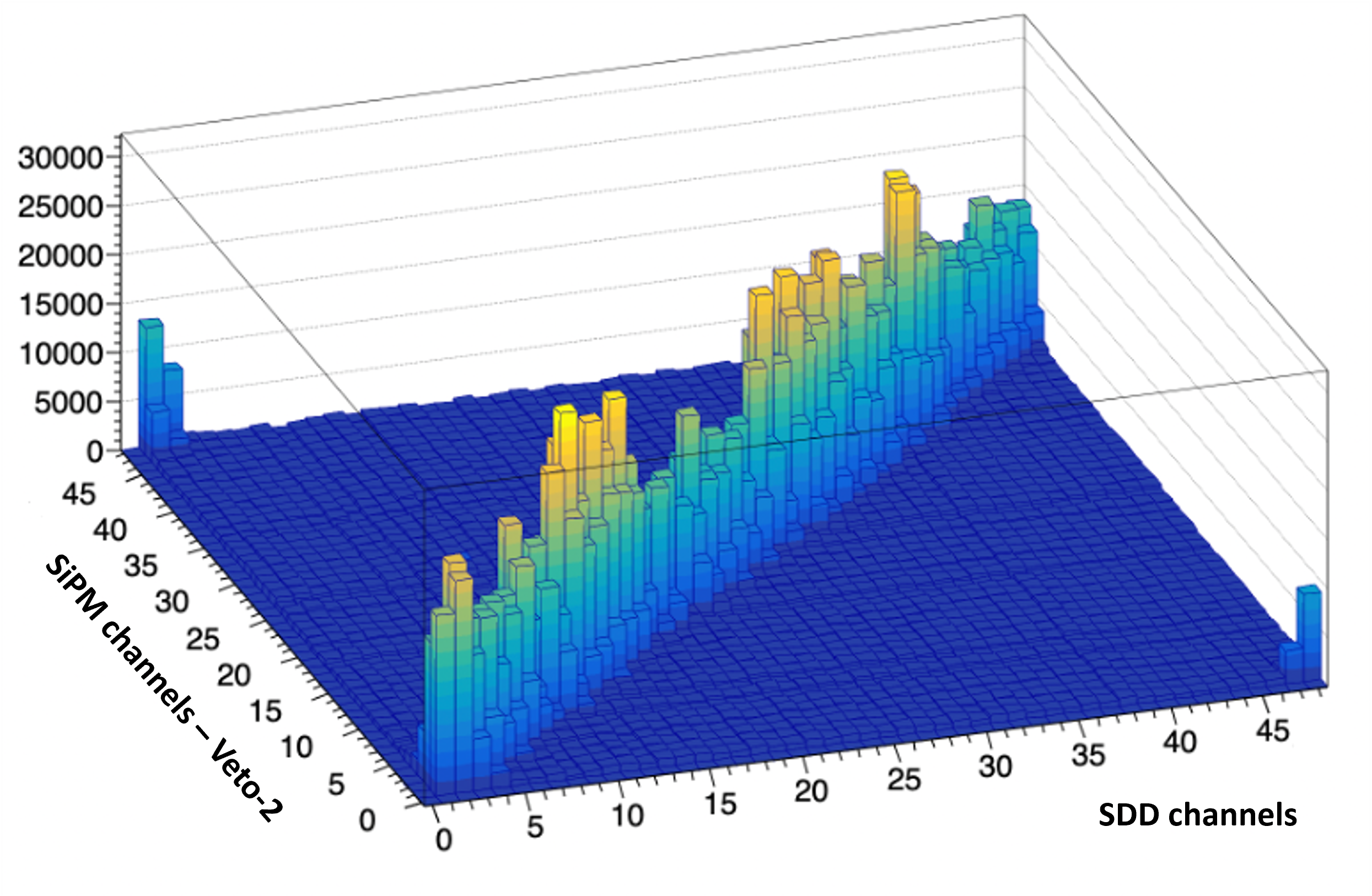}}
\caption{\small{Topological correlation between SDD and Veto-2 hits. Left: track of MIP passing the SDD channel and its corresponding Veto-2 scintillator. Right: The events correlation between the SDD vs SiPM channels from Veto-2 during the kaonic helium test run.}
\label{trackveto2}}
\end{figure}

The Veto-2 system has been tested and characterised at SMI, Vienna \cite{Marlene:2023}. The average time resolution of the Veto-2 system was (293 $\pm$ 45) ps, which allows to use the time information from the Veto-2 to distinguish between kaon stops in the target gas and stops in the solid structures of the setup, in analogy to the Veto-1 system. The intrinsic detection efficiency was found to be > 99$\%$, using cosmic rays. 
The full Veto-2 system was installed behind the SDDs and tested during the helium data taking period. To reject the signals produced by hadronic background, a coincidence of signals in the SDDs and in one of the corresponding Veto-2 detectors behind was required (see Fig.\ref{trackveto2}). To determine the efficiency, signals in the SDDs with energies > 20 keV were selected. They correspond to central hits of the SDDs by MIPs and therefore result in signals in the SDDs, the Veto-2, and the Veto-1. The efficiency was found to be 62 $\pm$ 1$\%$.

This efficiency of the Veto-2 system within the SIDDHARTA-2 apparatus is lower than its intrinsic detection efficiency (which is $>$99$\%$). This can be explained by an additional background component from beam-beam interactions present during the measurement, along with the hadronic background component. The Veto-2 system not only actively suppresses the hadronic background, but it also is instrumental for the understanding of the background composition and the optimisation of the machine and setup performances.

\subsubsection{The bottom kaon detector}\label{charged_section}

\begin{figure}[h]
\centering 
\mbox{\includegraphics[width=14 cm]{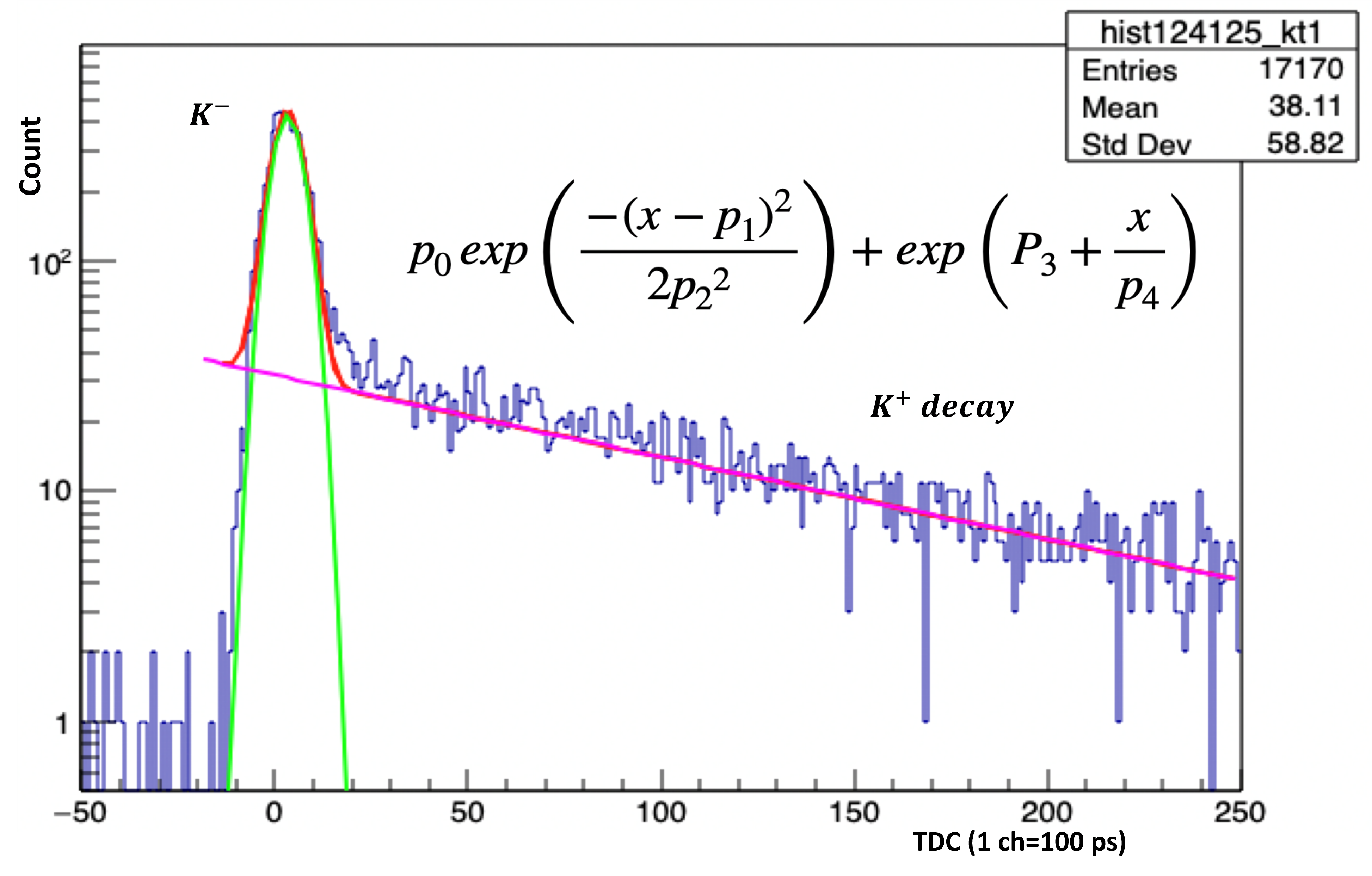}}
\caption{\small{Time spectrum of the bottom kaon detector. The main peak corresponds to the promptly emitted MIPs after the absorption of the $K^-$ in the detector, representing an event in which a $K^+$ is reaching the target cell. The exponential tail contains the good events where the $K^-$ was stopped in the gas target.}
\label{charged}}
\end{figure}

Complementary to the veto systems, a bottom kaon detector is installed below the lower scintillator of the kaon trigger system. It consists of a 50$\times$12$\times$5 mm$^3$ plastic scintillator covered by a 5 mm Teflon absorber on top, read out by PMTs. The purpose of this detector is the rejection of background produced by a $K^+$ reaching the target cell instead of the $K^-$. If a $K^-$ is stopped in the bottom detector, it is captured by the surrounding atoms and absorbed, leading to the prompt emission of MIPs within < 1 ns. Since the corresponding a $K^+$ is not captured into a kaonic atom, timing information can be used to distinguish between the promptly emitted MIPs from the bottom detector and the decay of the $K^+$with a mean lifetime of 12.4 ns. 
In contrast, the detection of a $K^+$ in the bottom kaon detector thus results in a delayed emission of MIPs. This difference in timing is used in the analysis to reject the events for which a positively charged kaon reached the target cell. Fig.~\ref{charged} shows the TDC spectrum of the bottom kaon detector and the fit function for the two components obtained during the helium run. The main peak represents the events in which a $K^+$ reached the target cell and a promptly emitted MIP was detected in the bottom detector. These events have to be discarded in the further analysis. The tail corresponds to the events used for further analysis, where a $K^-$ was stopped in the gas target.

\subsection{Trigger Logic and Data Acquisition System}\label{trigger_section}

\begin{figure}[htbp]
\centering 
\mbox{\includegraphics[width=14 cm]{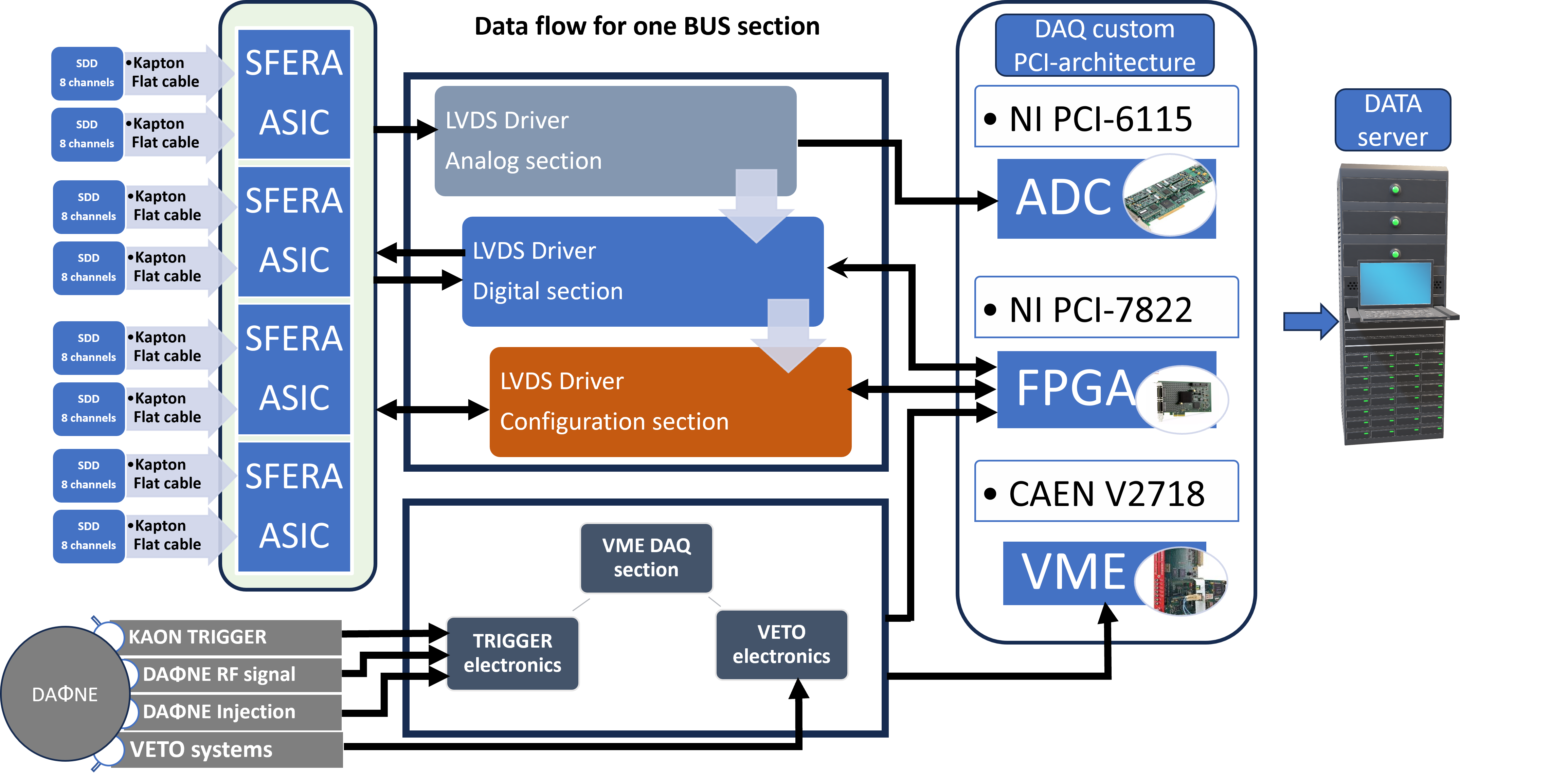}}
\caption{\small{Data flow chart for one BUS of the DAQ.}
\label{DAQ}}
\end{figure}

The analog output of one readout chip (8 channels SDD) is digitized by one of the 4 channel ADC converters of PCI-6115 board from National Instruments, with 12-bit resolution. Thus one ADC board manages the outputs of 4 readout chips with 64 SDD channels sharing one bus line (see Fig.\ref{DAQ}). All the 384 SDD channels are divided into 6 buses that are connected to 6 ADC boards. In this way, every SDD is uniquely identified by the BUS it belongs to and its serial number inside the BUS.\\
The kaon trigger scheme is defined as a coincidence between the upper and down scintillators of the kaon detector system. The analog signals from the PMTs are split into two outputs by the passive splitter; one is sent to the QDC (CAEN V792N) and the other to an Constant Fraction Discriminator (CFD) model ORTEC 935 with 200 MHz bandwidth. The outputs from the CFD are sent to a TDC (CAEN V775N) and to a Dual Mean Timer (CAEN N235). Coincidence signals between two outputs from the Mean Timers are sent to the gate of the QDC. Timing information for kaon event is stored using TDC with a resolution of 3.5~ps. The “RF/2” continuous pulse provided by DA$\Phi$NE radio frequency (3.68~GHz) and the time elapsed from the start to the rise time of the fast output pulse is used as the start signal for TDC recorded by each PMT (common start mode).

The kaon trigger and SDD detectors operate independently with separate logic without a conventional "master trigger" to start an event. In the SIDDHARTA-2 experiment, data coming from the two sectors are then combined at the final stage of the acquisition program. The DAQ is built on a PCI bus base, using National Instruments modules under the LabVIEW environment. The whole system is integrated by two PCIe-7822 Multifunction RIO boards, each one has an onboard Virtex-II 3M gate FPGA programmable with the LabVIEW FPGA package software.
The FPGA was compiled with the logic of the following functions: communication with interfaces of readout chip and kaon data; management of memory and timing difference between kaon signal and X-ray signal from SDD detectors; construction of event.\\ 
The signals of the veto systems are acquired by a VME (Versa Module Eurocard) TDC. A multi-hit TDC (CAEN V1190A) records the digital time over threshold (ToT) signals of the Veto-2. The ToT is defined as the timing difference between the leading and trailing edge of the digitized Veto-2 signal. It is proportional to the amplitude of the SiPM signal and therefore is a measure for the energy loss of the detected particle in the scintillator. At the stage of DAQ, the new scheme for a triple coincidence is defined using the kaon trigger information from kaon detector by this formula:
Coincidence signal = Kaon trigger signal $\cap$ SDDs signal
where the “$\cap$” means a coincidence of a Low Threshold (LT) real-time signal at SDD level and a kaon trigger signal generating a gate signal with $\sim$5~$\mu$s width. The X-ray events that satisfy the coincidence condition are tagged with a kaon trigger flag. The time difference between the X-ray event on SDD and kaon trigger (defined as “drift time”) is sampled by a 120~MHz onboard clock of the PCIe-7822 board. This timing difference information is included inside the SDD data event, and will be referred to as “drift time” for event selection of kaonic X-ray.

\section{Preliminary results with the SIDDHARTA-2 apparatus}\label{background_section}

\begin{figure}[h]
\centering 
\mbox{\includegraphics[width=12 cm]{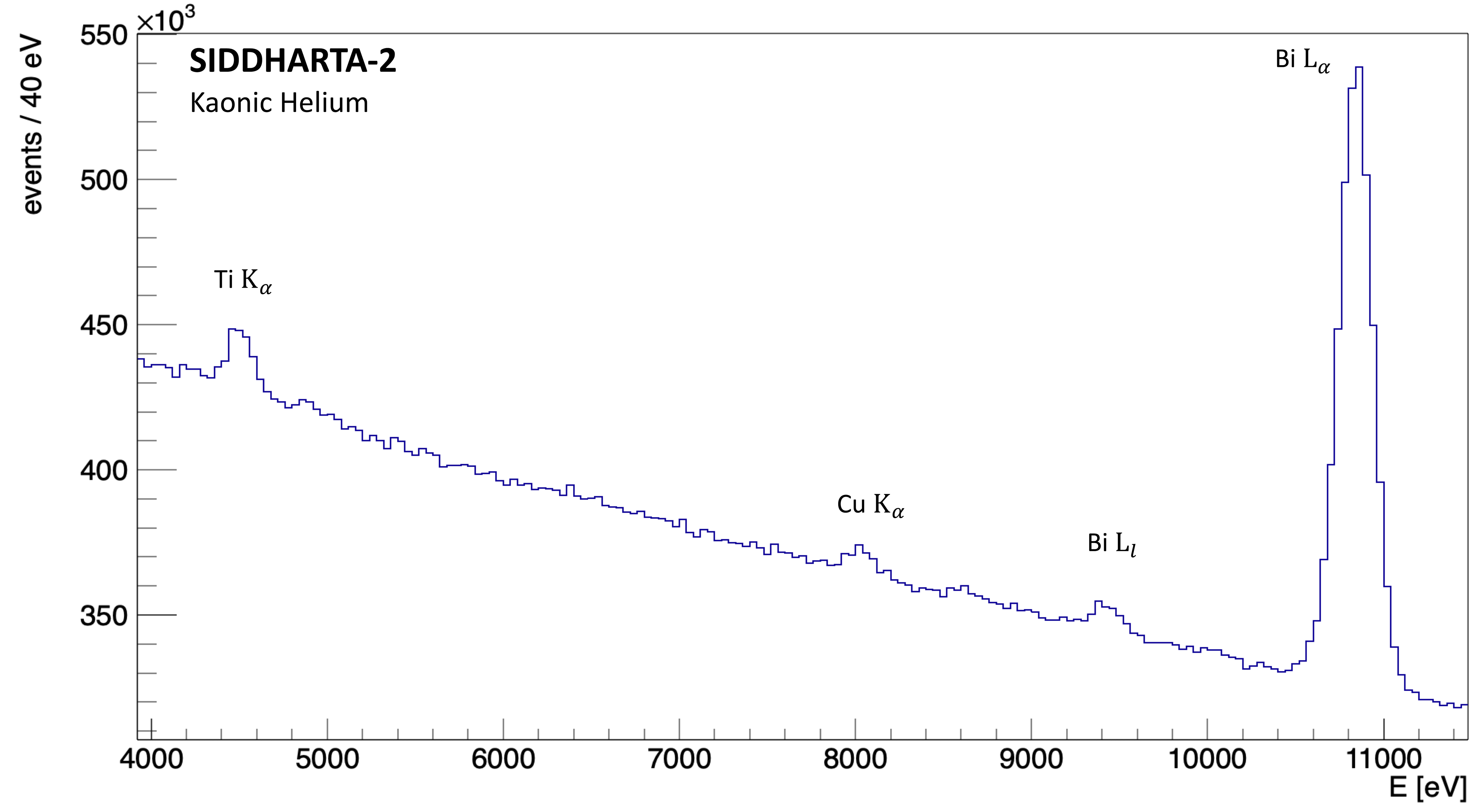}}
\caption{\small{Kaonic helium X-ray spectrum obtained in May 2023 (sample of 4.2~pb$^{-1}$) with no cuts applied. Fluorescence peaks from Ti, Cu and Bi are shown.}
\label{khe}}
\end{figure}

Before the kaonic deuterium measurement, a reduced version of the setup (SIDDHARTINO), with only 1/6 of the X-ray SDD, was installed in 2021 during the commissioning of the DA$\Phi$NE collider. The goal of this run was to optimize the performances of the machine, the dynamics of the beams, the performance of our experimental apparatus (trigger system, performance of the SDDs and optimization of the degrader to maximize the fraction of kaons in the gaseous target). All these optimizations were done via the measurement of the kaonic helium transitions to the 2p level. The SIDDHARTINO results \cite{Sirghi:2022wbj,Sirghi:2023scw} put even more stringent limit than the previous SIDDHARTA  kaonic helium-4 outcome \cite{SIDDHARTA:2012rsv,Bazzi:2014}, and exclude large shifts and widths.
After several improvements and optimizations of the SIDDHARTA-2 setup, described in the previous subsections, in the beginning of 2023 a new measurement of kaonic helium transitions was performed. The larger number of SDDs, and the varying background during the luminosity tuning, suggested to perform a technical run with helium before filling the target with deuterium. 
The inclusive energy spectrum acquired by the SDDs during the kaonic helium run, for a 4.2~pb$^{-1}$ sample is shown in Fig.\ref{khe}, before any cuts requirement. 

The high continuous backgrounds contribution below the peaks is mainly due to the asynchronous component of the machine background, while the Ti and Cu peaks originate from the calibration foils installed inside the target and activated by particles lost from the beams.  The Bi peak is produced by the activation of the alumina carrier behind the SDDs silicon wafer. 

\begin{figure}[h]
\centering 
\mbox{\includegraphics[width=12 cm]{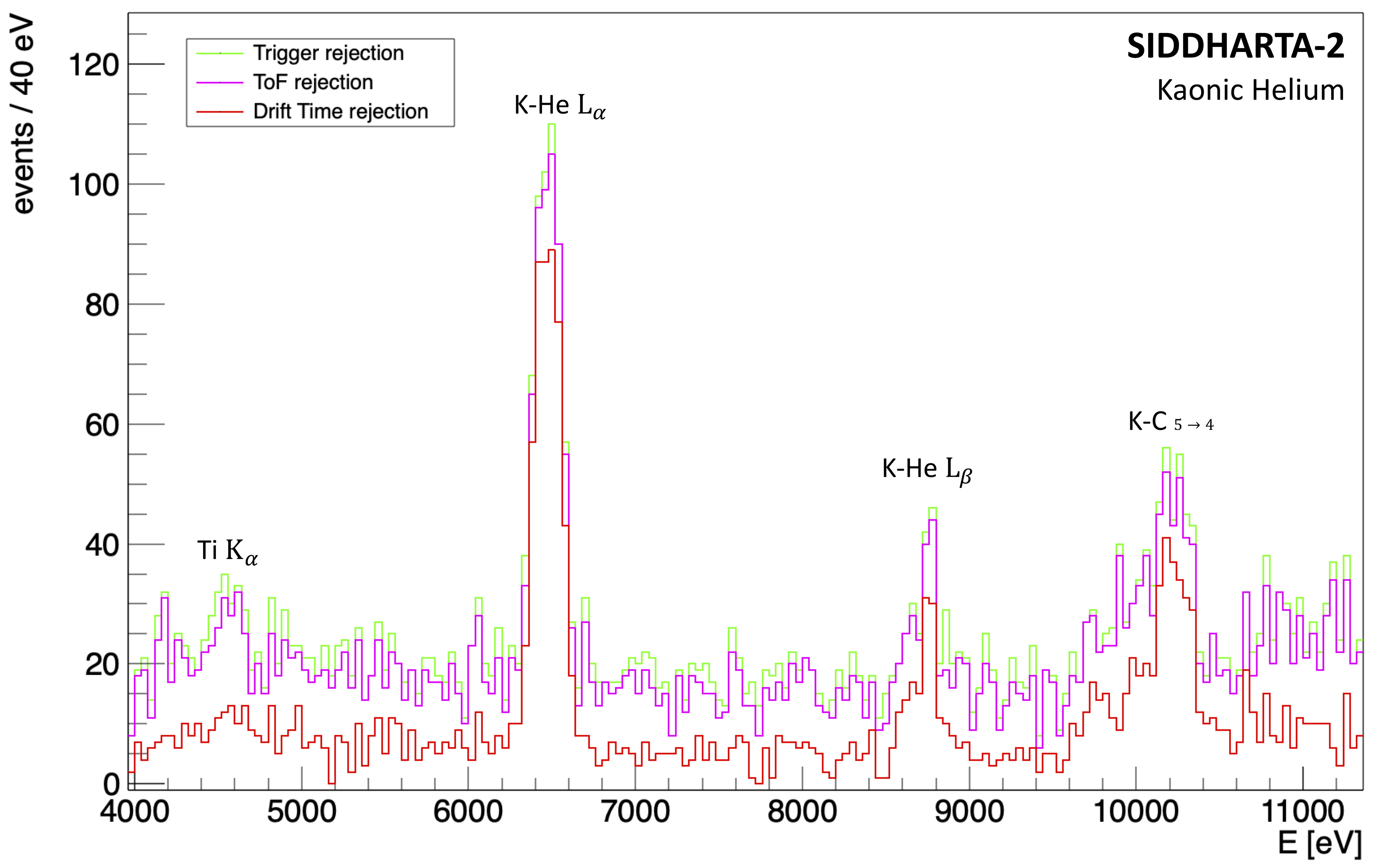}}
\caption{\small{Kaonic helium X-ray spectrum obtained after each step of background reduction: kaon trigger requirement (green), ToF selection (light magenta), and SDD drift time (red).}
\label{khe_cut}}
\end{figure}
To remove the asynchronous background, a first selection is applied using the kaon trigger information. Only the X-ray SDD signals within a 5 $\mu$s time window with respect to the trigger signal can pass the trigger selection. The time window width was tuned to enable the front-end electronics to process and acquire the signals.
Using the ToF information from the kaon trigger, a part of the remaining synchronous background due to the MIPs produced in electromagnetic showers can be eliminated.   
Moreover, the SDD drift time information registered by the DAQ was used to further reduce the asynchronous background. 
The effects of all these selections are presented in Fig.\ref{khe_cut}, where the spectra with the kaon trigger rejection (green), the further selection of the kaons from ToF (magenta) and the SDD drift time requirements (red), are shown. 
The results confirm those obtained during the SIDDHARTINO run \cite{Sirghi:2022wbj}.

\section{Conclusions}

The DA$\Phi$NE collider at INFN-LNF stands out as the optimal source of kaons for conducting precise measurements on kaonic atoms. The installation of the SIDDHARTA-2 apparatus has successfully been concluded during the commissioning phase of the collider. The setup with all improvements and optimization has proven its functionality and adequacy for the challenging kaonic atom spectroscopy measurements \cite{Sirghi:2022wbj,Sgaramella:2023orc,Sgaramella_2023}. The SIDDHARTA-2 experiment is currently taking data for the measurement of kaonic deuterium which was initiated in the spring of 2023, aiming to collect 800~pb$^{-1}$ of data. The experiment aims to provide a kaonic deuterium measurement at the same level of precision as the kaonic hydrogen one. 
Further experiments have been proposed at DA$\Phi$NE, following the completion of the kaonic deuterium project\cite{Curceanu2021,Curceanu2021b}.

\section*{Acknowledgments}

We thank C. Capoccia from LNF-INFN and H. Schneider, L. Stohwasser, and D. Pristauz-Telsnigg from Stefan-Meyer-Institut for their fundamental contribution in designing and building the SIDDHARTA-2 setup. We thank as well the DA$\Phi$NE staff for the excellent working conditions and permanent support. Part of this work was supported by the Austrian Science Fund (FWF): [P24756-N20 and P33037-N]; the Croatian Science Foundation under the project IP-2018-01-8570; EU STRONG-2020 project (grant agreement No.~824093); the EU Horizon 2020 project under the MSCA G.A. 754496; the EXOTICA project of the Minstero degli Affari Esteri e della Cooperazione Internazionale, PO22MO03; the Polish Ministry of Science and Higher Education grant No. 7150/E-338/M/2018; the Polish National Agency for Academic Exchange (grant No. PPN/BIT/2021/1/00037); the SciMat and qLife Priority Research Areas budget under the program Excellence Initiative – Research University at the Jagiellonian University and the Foundational Questions Institute and Fetzer Franklin Fund, a donor-advised fund of Silicon Valley Community Foundation (Grant No. FQXi-RFP-CPW-2008).

\bibliographystyle{elsarticle-num} 
\bibliography{cas-refs}

\begin{thebibliography}{10}
\expandafter\ifx\csname url\endcsname\relax
  \def\url#1{\texttt{#1}}\fi
\expandafter\ifx\csname urlprefix\endcsname\relax\def\urlprefix{URL }\fi
\expandafter\ifx\csname href\endcsname\relax
  \def\href#1#2{#2} \def\path#1{#1}\fi

\bibitem{RevModPhys.91.025006}
C.~Curceanu, C.~Guaraldo, M.~Iliescu, M.~Cargnelli, R.~Hayano, J.~Marton, J.~Zmeskal, T.~Ishiwatari, M.~Iwasaki, S.~Okada, D.~L. Sirghi, H.~Tatsuno, \href{https://link.aps.org/doi/10.1103/RevModPhys.91.025006}{The modern era of light kaonic atom experiments}, Rev. Mod. Phys. 91 (2019) 025006.
\newblock \href {https://doi.org/10.1103/RevModPhys.91.025006} {\path{doi:10.1103/RevModPhys.91.025006}}.
\newline\urlprefix\url{https://link.aps.org/doi/10.1103/RevModPhys.91.025006}

\bibitem{KpX:1997}
M.~Iwasaki, R.~S. Hayano, T.~M. Ito, S.~N. Nakamura, T.~P. Terada, D.~R. Gill, L.~Lee, A.~Olin, M.~Salomon, S.~Yen, K.~Bartlett, G.~A. Beer, G.~Mason, G.~Trayling, H.~Outa, T.~Taniguchi, Y.~Yamashita, R.~Seki, \href{https://link.aps.org/doi/10.1103/PhysRevLett.78.3067}{Observation of kaonic hydrogen ${K}_{\ensuremath{\alpha}}$ x rays}, Phys. Rev. Lett. 78 (1997) 3067--3069.
\newblock \href {https://doi.org/10.1103/PhysRevLett.78.3067} {\path{doi:10.1103/PhysRevLett.78.3067}}.
\newline\urlprefix\url{https://link.aps.org/doi/10.1103/PhysRevLett.78.3067}

\bibitem{Bazzi:2011}
M.~Bazzi, et~al., {A new measurement of kaonic hydrogen X-rays}, Phys. Lett. B 704~(3) (2011) 113.
\newblock \href {https://doi.org/10.1016/j.physletb.2011.09.011} {\path{doi:10.1016/j.physletb.2011.09.011}}.

\bibitem{BAZZI201369}
M.~Bazzi, et~al., Preliminary study of kaonic deuterium x-rays by the siddharta experiment at da$\phi$ne, Nuclear Physics A 907 (2013) 69--77.
\newblock \href {https://doi.org/https://doi.org/10.1016/j.nuclphysa.2013.03.001} {\path{doi:https://doi.org/10.1016/j.nuclphysa.2013.03.001}}.

\bibitem{Gal:2007}
A.~Gal, et~al., {On the Scattering Length of the $K^-$d System}, Int. J. Mod. Phys. A 221 (2007) 226–233.
\newblock \href {https://doi.org/10.1142/S0217751X07035379} {\path{doi:10.1142/S0217751X07035379}}.

\bibitem{Meisner:2011}
M.~Doring, U.~Meißner, et~al., {Kaon–nucleon scattering lengths from kaonic deuterium experiments revisited}, Phys. Lett. B 704 (2011) 663–666.
\newblock \href {https://doi.org/10.1016/j.physletb.2011.09.099} {\path{doi:10.1016/j.physletb.2011.09.099}}.

\bibitem{Shevchenko:2012}
N.~Shevchenko, et~al., {Near-threshold $K^-$d scattering and properties of kaonic deuterium}, Nucl. Phys. A 890-891 (2012) 50–61.
\newblock \href {https://doi.org/10.1016/j.nuclphysa.2012.07.010} {\path{doi:10.1016/j.nuclphysa.2012.07.010}}.

\bibitem{Mizutani:2013}
T.~Mizutani, et~al., {Faddeev-chiral unitary approach to the $K^-$d scattering length}, Phys. Rev. C 87 (2013) 035.
\newblock \href {https://doi.org/10.1103/PhysRevC.87.035201} {\path{doi:10.1103/PhysRevC.87.035201}}.

\bibitem{Liu:2020}
Z.-W. Liu, et~al., {Kaonic hydrogen and deuterium in Hamiltonian effective field theory}, Phys. Lett. B 808 (2020) 135.
\newblock \href {https://doi.org/10.1016/j.physletb.2020.135652} {\path{doi:10.1016/j.physletb.2020.135652}}.

\bibitem{Zyla:2020zbs}
P.~Zyla, et~al., {Review of Particle Physics}, PTEP 2020~(8) (2020) 083C01.
\newblock \href {https://doi.org/10.1093/ptep/ptaa104} {\path{doi:10.1093/ptep/ptaa104}}.

\bibitem{Zobov:2010zza}
M.~Zobov, et~al., {Test of crab-waist collisions at DA\ensuremath{\Phi}NE $\Phi$ factory}, Phys. Rev. Lett. 104 (2010) 174801.
\newblock \href {https://doi.org/10.1103/PhysRevLett.104.174801} {\path{doi:10.1103/PhysRevLett.104.174801}}.

\bibitem{Sirghi:2022wbj}
D.~Sirghi, et~al., {A new kaonic helium measurement in gas by SIDDHARTINO at the DA\ensuremath{\Phi}NE collider}, J. Phys. G: Nucl. Part. Phys. 49~(5) (2022) 055106.

\bibitem{jpet1}
P.~Moskal, et~al., {Positronium imaging with the novel multiphoton PET scanner}, Sci. Adv. 7 (2021) eabh4394.
\newblock \href {https://doi.org/10.1126/sciadv.abh4394} {\path{doi:10.1126/sciadv.abh4394}}.

\bibitem{jpet2}
P.~Moskal, et~al., {Testing CPT symmetry in ortho-positronium decays with positronium annihilation tomography}, Nat. Commun. 12 (2021) 5658.
\newblock \href {https://doi.org/10.1038/s41467-021-25905-9} {\path{doi:10.1038/s41467-021-25905-9}}.

\bibitem{jpet3}
P.~Moskal, et~al., {Test of a single module of the J-PET scanner based on plastic scintillators}, Nucl. Instrum. Methods Phys. Res. A 764 (2014) 317.
\newblock \href {https://doi.org/10.1016/j.nima.2014.07.052} {\path{doi:10.1016/j.nima.2014.07.052}}.

\bibitem{Skurzok:2020phi}
M.~Skurzok, et~al., {Characterization of the SIDDHARTA-2 luminosity monitor}, JINST 15~(10) (2020) P10010.
\newblock \href {http://arxiv.org/abs/2008.05472} {\path{arXiv:2008.05472}}, \href {https://doi.org/10.1088/1748-0221/15/10/P10010} {\path{doi:10.1088/1748-0221/15/10/P10010}}.

\bibitem{koike1996}
T.~Koike, T.~Harada, Y.~Akaishi, \href{https://link.aps.org/doi/10.1103/PhysRevC.53.79}{Cascade calculation of ${\mathit{k}}^{\mathrm{\ensuremath{-}}}$-p and ${\mathit{k}}^{\mathrm{\ensuremath{-}}}$-d atoms}, Phys. Rev. C 53 (1996) 79--87.
\newblock \href {https://doi.org/10.1103/PhysRevC.53.79} {\path{doi:10.1103/PhysRevC.53.79}}.
\newline\urlprefix\url{https://link.aps.org/doi/10.1103/PhysRevC.53.79}

\bibitem{Gatti:1984}
E.~Gatti, P.~Rehak, {Semiconductor drift chamber — An application of a novel charge transport scheme}, Nucl. Instrum. Methods Phys. Res. 225 (1984) 608–614.
\newblock \href {https://doi.org/10.1016/0167-5087(84)90113-3} {\path{doi:10.1016/0167-5087(84)90113-3}}.

\bibitem{cube:2010}
L.~Bombelli, C.~Fiorini, T.~Frizzi, R.~Nava, A.~Greppi, A.~Longoni, Low-noise cmos charge preamplifier for x-ray spectroscopy detectors, in: IEEE Nuclear Science Symposuim "I\&" Medical Imaging Conference, 2010, pp. 135--138.
\newblock \href {https://doi.org/10.1109/NSSMIC.2010.5873732} {\path{doi:10.1109/NSSMIC.2010.5873732}}.

\bibitem{Quaglia:2016uox}
R.~Quaglia, F.~Schembari, G.~Bellotti, A.~D. Butt, C.~Fiorini, L.~Bombelli, G.~Giacomini, F.~Ficorella, C.~Piemonte, N.~Zorzi, {Development of arrays of Silicon Drift Detectors and readout ASIC for the SIDDHARTA experiment}, Nucl. Instrum. Meth. A 824 (2016) 449--451.
\newblock \href {https://doi.org/10.1016/j.nima.2015.08.079} {\path{doi:10.1016/j.nima.2015.08.079}}.

\bibitem{Schembari:2016IEE}
F.~Schembari, et~al., {SFERA: An Integrated Circuit for the Readout of X and $\gamma$-Ray Detectors.}, IEEE Trans. Nucl. Sci. 63 (2016) 1797.
\newblock \href {https://doi.org/10.1109/TNS.2016.2565200} {\path{doi:10.1109/TNS.2016.2565200}}.

\bibitem{Miliucci:2019mdpi}
M.~Miliucci, et~al., {Energy Response of Silicon Drift Detectors for Kaonic Atom Precision Measurements}, Condensed Matter 4 (2019) 31.
\newblock \href {https://doi.org/10.3390/condmat4010031} {\path{doi:10.3390/condmat4010031}}.

\bibitem{Miliucci:2021wbj}
M.~Miliucci, et~al., {Silicon drift detectors system for high-precision light kaonic atoms spectroscopy}, Measur. Sci. Tech. 32~(9) (2021) 095501.
\newblock \href {https://doi.org/10.1088/1361-6501/abeea9} {\path{doi:10.1088/1361-6501/abeea9}}.

\bibitem{Bazzi2013}
M.~Bazzi, C.~Berucci, C.~Curceanu, A.~d'Uffizi, M.~Iliescu, E.~Sbardella, A.~Scordo, H.~Shi, F.~Sirghi, H.~Tatsuno, I.~Tucakovic, Characterization of the siddharta-2 second level trigger detector prototype based on scintillators coupled to a prism reflector light guide, Journal of Instrumentation 8~(11) (2013) T11003.
\newblock \href {https://doi.org/10.1088/1748-0221/8/11/T11003} {\path{doi:10.1088/1748-0221/8/11/T11003}}.

\bibitem{Marlene:2023}
M.~Tüchler, {Probing the Strong Interaction with Kaonic Atom X-Ray Measurements at Low Energies}, Vienna (2023) Ph.D Thesis.

\bibitem{Marlene;2018}
M.~T\"uchler, et~al., {A charged particle veto detector for kaonic deuterium measurements at DA$\Phi$NE}, Journal of Physics: Conference Series 1138~(1) (2018) 012012.
\newblock \href {https://doi.org/10.1088/1742-6596/1138/1/012012} {\path{doi:10.1088/1742-6596/1138/1/012012}}.

\bibitem{Sirghi:2023scw}
D.~L. Sirghi, et~al., {New measurements of kaonic helium-4 L-series X-rays yields in gas with the SIDDHARTINO setup}, Nucl. Phys. A 1029 (2023) 122567.
\newblock \href {https://doi.org/10.1016/j.nuclphysa.2022.122567} {\path{doi:10.1016/j.nuclphysa.2022.122567}}.

\bibitem{SIDDHARTA:2012rsv}
M.~Bazzi, et~al., {Measurements of the strong-interaction widths of the kaonic 3He and 4He 2p levels}, Phys. Lett. B 714 (2012) 40--43.
\newblock \href {http://arxiv.org/abs/1205.0640} {\path{arXiv:1205.0640}}, \href {https://doi.org/10.1016/j.physletb.2012.06.071} {\path{doi:10.1016/j.physletb.2012.06.071}}.

\bibitem{Bazzi:2014}
M.~Bazzi, et~al., {L-series X-ray yields of kaonic $^3He$ and $^4He$ atoms in gaseous targets}, Eur. Phys. J. A 50 (2014) 91.
\newblock \href {https://doi.org/10.1140/epja/i2014-14091-0} {\path{doi:10.1140/epja/i2014-14091-0}}.

\bibitem{Sgaramella:2023orc}
F.~Sgaramella, et~al., {Measurements of high-n transitions in intermediate mass kaonic atoms by SIDDHARTA-2 at DA$\mathrm {\Phi }$NE}, Eur. Phys. J. A 59~(3) (2023) 56.
\newblock \href {http://arxiv.org/abs/2304.11352} {\path{arXiv:2304.11352}}, \href {https://doi.org/10.1140/epja/s10050-023-00976-y} {\path{doi:10.1140/epja/s10050-023-00976-y}}.

\bibitem{Sgaramella_2023}
F.~Sgaramella, et~al., \href{https://dx.doi.org/10.1088/1742-6596/2446/1/012023}{Kaonic atoms measurements with siddharta-2}, Journal of Physics: Conference Series 2446~(1) (2023) 012023.
\newblock \href {https://doi.org/10.1088/1742-6596/2446/1/012023} {\path{doi:10.1088/1742-6596/2446/1/012023}}.
\newline\urlprefix\url{https://dx.doi.org/10.1088/1742-6596/2446/1/012023}

\bibitem{Curceanu2021}
C.~Curceanu, et~al., {Kaonic Atoms Measurements at DA\ensuremath{\Phi}NE: SIDDHARTA-2 and Future Perspectives}, Few-Body Syst., 62 (2021) 83.
\newblock \href {https://doi.org/10.1007/s00601-021-01666-5} {\path{doi:10.1007/s00601-021-01666-5}}.

\bibitem{Curceanu2021b}
C.~Curceanu, et~al., {Fundamental physics at the strangeness frontier at DA\ensuremath{\Phi}NE. Outline of a proposal for future measurements} (2021).
\newblock \href {https://doi.org/arXiv:2104.06076v2} {\path{doi:arXiv:2104.06076v2}}.

\end{thebibliography}

\end{document}